\documentclass[
	pra,
	aps,
	reprint,
	amsmath,amssymb,
	a4paper,
	superscriptaddress,
	10pt,
    nofootinbib
]{revtex4-2}
\usepackage{helvet}
\usepackage{graphicx}
\usepackage{physics}
\graphicspath{{fig/}}
\usepackage[dvipsnames,table,xcdraw]{xcolor}
\definecolor{rmpblue}{HTML}{2e3092}
\usepackage[
	colorlinks=true,
	citecolor=rmpblue,
	linkcolor=rmpblue,
	filecolor=magenta,
	urlcolor=rmpblue,
]{hyperref}
\usepackage[
	colorlinks=true
]{hyperref}
\usepackage{tikz}

\newcommand{\qquote}[1]{``#1''}

\newcommand{\affilANU}{Nonlinear Physics Center, Research School of Physics, Australian National University, Canberra ACT 2601, Australia}

\newcommand{\affilShubnikov}{ Shubnikov Institute of Crystallography, NRC ``Kurchatov Institute'', Moscow 119333, Russia}
\newcommand{\affilMEPhI}{National Research Nuclear University MEPhI (Moscow Engineering Physics Institute), Moscow 115409, Russia}
\newcommand{\affilBombay}{Department of Physics, Indian Institute of Technology Bombay, Mumbai 400076, India}

\newcommand{\iu}{\mathrm{i}\mkern1mu}
\newcommand{\eu}{\mathrm{e}\mkern1mu}

\newcommand{\ort}[1]{\boldsymbol{\mathbf{\hat{#1}}}}

\begin{document}
	
\title{Maximal optical chirality via mode coupling in bilayer metasurfaces}
\author{Brijesh Kumar}
\email{brijeshgangwar93@gmail.com}
\altaffiliation{These authors contributed equally.}
\affiliation{\affilANU}
\affiliation{\affilBombay}

\author{Ivan Toftul}
\email{toftul.ivan@gmail.com}
\altaffiliation{These authors contributed equally.}
\affiliation{\affilANU}

\author{Anshuman Kumar}
\affiliation{\affilBombay}

\author{Maxim Gorkunov}
\affiliation{\affilShubnikov}
\affiliation{\affilMEPhI}

\author{Yuri Kivshar}
\email{yuri.kivshar@anu.edu.au}
\affiliation{\affilANU}
	
\date{\today}
	
\begin{abstract}
Recent advances in the physics of resonant optical metasurfaces allowed to realize the so-called {\it maximum chirality} of planar structures by engineering their geometric parameters. Here we employ bilayer membrane metasurfaces with a square lattice of rotated C$_4$-symmetric holes and uncover very different scenarios of chirality maximization by virtue of strong coupling of photonic eigenmodes of the membranes supplemented by  smart engineering of dissipation losses. Our findings substantially expand the class of planar maximally chiral resonant surfaces feasible for widespread nanolithography techniques desired for metaphotonic applications in chiral sensing, chiral light emission, detection and polarization conversion. 
\end{abstract}
   
\maketitle

\section{Introduction}

Chirality is an important concept in physics applied to objects of any sizes and nature that are not identical with their mirror images~\cite{Barron2012Chirality,Caloz2020IEEE1,Caloz2020IEEE2}. Such objects are usually manifested in the left and right enantiomers which may possess different physical properties ~\cite{Deng2024NanoPhot,Sharma2009Science}. 
Optical chirality is a more specific characteristics of  electromagnetic properties of an object responding differently to left--circularly polarized (LCP) or right-circularly polarized (RCP) light. The key parameter quantifying optical chirality is {\it circular dichroism} (CD)~\cite{Woody1995book,berova2000circularDichroism,rodger1997circularDichroism}.

Chiral response is usually weak in natural materials, and the CD values are very low. Recently, chiral metasurfaces have been explored extensively for achieving remarkable values of CD by engineering artificial optical chirality which can be made much stronger than natural optical chirality. In spite of many advanced designs with plasmonics metasurfaces~\cite{Yu2016plasmonicMSchiral,Wang2019plasmonicMSchiral,Ouyang:2018:optexp:plasmonic,Khaliq2023:AOM:Plasmonic:and:dielectric}, it was understood very recently that resonant dielectric nanostructures can provide even higher values of CD by engineering  the structure of photonic eigenmodes~\cite{Khaliq2023:AOM:Plasmonic:and:dielectric,Solomon2019:ACSPhot:dielectricchiral,Wu:23:dielectric:chiral,Beutel2021:alldielectric,Tanaka:acsnano:2020:dielctric}. More specifically, one can select a metasurface mode of a certain helicity by employing the concept of “chiral quasi-bound state in the continuum” (chiral quasi-BIC)~\cite{Gorkunov2020PRL,Gorkunov2021:BIC,chen2023:BICchiral,overvig2021chiral}
and realize the so-called {\it maximum optical chirality}, when an optical structure is transparent to the waves of one helicity but reflects or absorbs waves of the opposite helicity.

Because many optical planar structures are fabricated on transparent substrates that break out-of-plane mirror symmetry, an important question is how a substrate can affect optical chirality. Recently, Gorkunov et al. \cite{Gorkunov2025AOM} revealed that low-refractive-index substrates can induce
up to maximum intrinsic optical chirality in otherwise achiral metastructures; this effect originates from engineered coupling of twisted photonic modes of different parity. 

Here, we extend the idea of optical chirality induced by symmetry-breaking mode coupling onto bilayer dielectric membrane metasurfaces (MM). As model systems, we consider MMs built as square lattices of $\mathrm{C}_{4h}$ symmetric holes, which inherently support degenerate resonant modes with well-defined spatial parity. The presence of a mirror symmetry plane renders such systems geometrically achiral, which strictly prohibits circular dichroism due to symmetry constraints regardless of  presence of optical resonances. To enable optical chirality,  in the absence of the LCP/RCP cross-conversion, we add a second MM layer, thereby reducing the symmetry from $\mathrm{C}_{4h}$ to $\mathrm{C}_4$, which preserves fourfold rotational symmetry but breaks the mirror plane, making the system geometrically chiral~\cite{Sinev2024arXiv}. This symmetry reduction gives rise to {\it chiral hybridization} of photonic modes.
However, due to reciprocity restrictions, no CD can be observed in rotation symmetric structures at normal incidence in the absence of dissipation: the transmission CD remains absent even after the geometric chirality is introduced. 

Incorporating controlled losses, particularly in the spectral ranges where the chiral mode hybridization is strong, 
one can break this balance and enable non-zero transmission CD. In these regimes, the interference between nearly-degenerate chiral modes becomes sensitive to loss channels, leading to asymmetric dissipation rates for the two circular polarizations. As a result, an MM acquires significant CD but the particular scenarios can be remarkably different: the CD sign and magnitude are strongly linked to the parity and symmetry character of the coupled resonant modes.

Our findings highlight how non-Hermiticity, when combined with precisely engineered symmetry breaking, can serve as a powerful tool to realize maximally chiral metasurfaces. In particular, we demonstrate that strong mode coupling, induced by the bilayer membrane geometry, provides an ideal platform to tune and enhance CD through the interplay of interference and loss. This approach offers a robust and scalable strategy for designing planar chiral photonic structures with tunable and enhanced chiroptical responses.

\begin{figure*}[t]
	\centering
	\begin{tikzpicture}
		\node[anchor=north west,inner sep=0] at (0,0)
		{\includegraphics[width=0.9\linewidth]{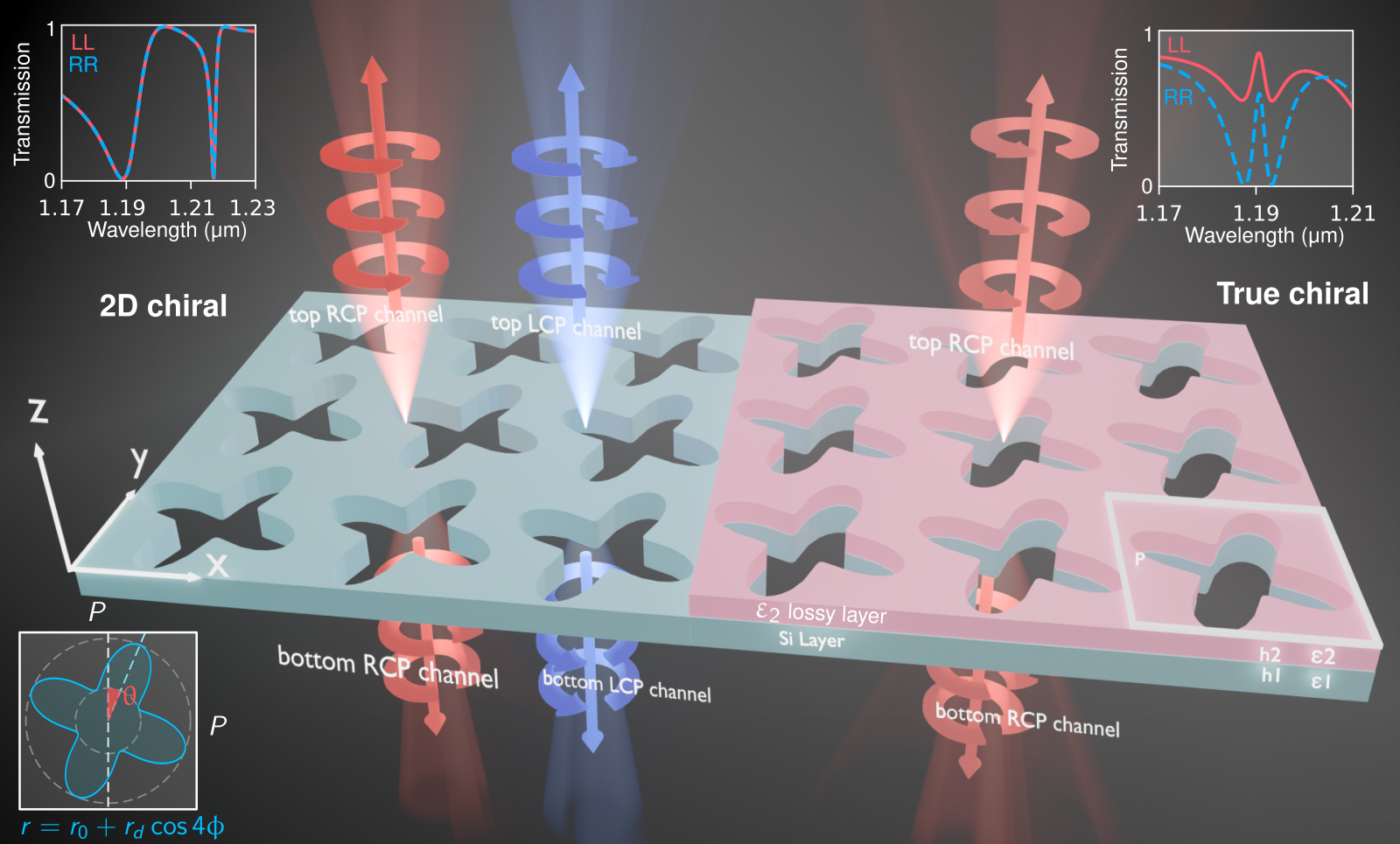}};
	\end{tikzpicture}
	\caption{\textbf{Concept of chiral bilayer membrane metasurface.} The schematic illustrates the transition from the single-layer achiral to the bilayer chiral structure. Achiral eigenmodes of the single layer structure give rise to achiral transmission resonances (see the top left inset), while chiral modes of the bilayer structure produce strong transmission CD, as illustrated in the top right inset. The bottom inset shows the designed unit cell with $\mathrm{C}_4$ rotational symmetry with the dashed lines indicating the inner and outer radii, $r_0 - r_d$ and $r_0 + r_d$ of the four-petal hole respectively.}
	\label{fig:concept}
\end{figure*}

\section{Breaking mirror symmetry}
\label{sec:Design of the structure}

\subsection{Main principle}
\label{sec:principle}

We consider a $\mathrm{C}_4$-symmetric metasurface build as a membrane perforated with a square lattice of four-petal holes schematically shown in  Fig.~\ref{fig:concept}. The membrane material is set to emulate silicon (dielectric permittivity $\varepsilon_r = 12$) with the frequency dispersion neglected for clarity. The host material is assumed to be air with a unit relative permittivity. In the following, the membrane thickness is varied between $0.4 P$ to $0.6 P$, where $P$ is the square lattice period. The hole shape is parameterized in polar coordinates as a perturbed circle:  $r(\phi) = r_0 + r_d\cos{(4\phi)}$, where $r_0 = 0.35 P$ and $r_d = 0.15P$. To break the in-plane mirror symmetry and induce two-dimensional (2D) chirality, the hole shape is rotated by $\pi/8$ with respect to the lattice.  

The $\mathrm{C}_4$ symmetry inherently prevents polarization conversion during transmission of normally incident LCP and RCP waves, while the reflections are fully cross-polarized: LCP waves are reflected as RCP and vice versa. As the out-of-plane mirror symmetry also implies identical co-polarized transmission of LCP and RCP waves, such 2D chiral metasurface totally lacks optical chirality (see more in Appendix~\ref{app:CDzero} and in Refs~\cite{Gorkunov2024, Shalin2023,Koshelev2024JOPT}).

To transform the 2D chiral structure into truly chiral one, we break the out-of plane mirror symmetry by introducing \textit{an adjacent layer} (with a dielectric permittivity $\varepsilon_r = 3$) perforated with the same lattice of holes. In the following, we refer to the single layer configuration as a 2D chiral MM while the bilayer structure is referred to as a truly chiral MM. 

Chiral symmetry breaking by the second layer enables optical chirality, which, due to the remaining $\mathrm{C}_4$ symmetry, is manifested by the co-polarized circular dichroism~\cite{Toftul2024PRL}:
\begin{equation}
	\mathrm{CD}_{\mathrm{co}} = \frac{|t_{\text{RR}}|^2 - |t_{\text{LL}}|^2}{|t_{\text{RR}}|^2 + |t_{\text{LL}}|^2},
	\label{eq:CDco}
\end{equation}
where $t_{\text{RR}}$ and $t_{\text{LL}}$ are the complex transmission coefficients in the circular polarization basis, with the first and last indexes denoting the output and input polarizations, respectively. Due to the Lorentz reciprocity, the transmission coefficients are the same for light incident on either metasurface side. We stress that for absent geometrical chirality automatically  $\mathrm{CD}_{\mathrm{co}} = 0$~\cite{Koshelev2024JOPT,Shalin2023}, so here optical chirality --- a generally different optical response to RCP and LCP light --- can be used to quantify geometric chirality. Note that such a correlation is not always present, as geometrically achiral structures of lower rotation symmetry can exhibit strong optical chirality~\cite{Semnani2020, Voronin2022}.

To understand the origin and mechanisms of optical chirality, it is very useful to analyze the MM photonic eigenmodes also known as resonant states or quasi-normal modes \cite{Lalanne2018}. Defined as solutions of source-free Maxwell's equations, such modes underpin all optical resonances and, in particular, the observable characteristics of optical chirality can be traced down to the dissymmetry of excitation and irradiation of specific chiral eigenmodes \cite{Kondratov2016, Gorkunov2020PRL, Toftul2024PRL}. To quantify the contribution of an eigenmode to the chiral transmission dissymmetry, it is convenient to introduce \textit{the modal circular dichroism} as:
\begin{equation}
	\mathrm{CD}_{\text{mode}, n} = \frac{|m_{n\mathrm{R}} m^{\prime}_{n \mathrm{R}}|^2 - |m_{n\mathrm{L}} m^{\prime}_{n\mathrm{L}}|^2}{|m_{n\mathrm{R}} m^{\prime}_{n \mathrm{R}}|^2 + |m_{n\mathrm{L}} m^{\prime}_{n\mathrm{L}}|^2}.
	\label{eq:CD_mode}
\end{equation}
which is built from the parameters $m_{n \mathrm{R}}$ and $m_{n \mathrm{L}}$ describing the coupling of an $n$-th mode to propagating in the $-z$ direction RCP and LCP waves respectively:
\begin{align}
	\begin{split}
		m_{n \mathrm{R}} &= A_n \int \limits_{\mathbb{V}_{\text{MM}}}  \left[\varepsilon(
		\vb{r})  - 1\right]\mathbf{E}_n(\mathbf{r}) \cdot \ort{e}_{+} \eu^{-\iu \Omega_n z/c} \, \dd V, \\
		m_{n \mathrm{L}} &= A_n \int \limits_{\mathbb{V}_{\text{MM}}}  \left[\varepsilon(
		\vb{r})  - 1\right]\mathbf{E}_n(\mathbf{r}) \cdot \ort{e}_{-} \eu^{-\iu \Omega_n z/c} \, \dd V,
	\end{split}
	\label{eq:m}
\end{align}
while those for waves propagating in the opposite $+z$ direction read as:
\begin{align}
	\begin{split}
		m^{\prime}_{n \mathrm{R}} &= A_n \int \limits_{\mathbb{V}_{\text{MM}}}  \left[\varepsilon(
		\vb{r})  - 1\right]\mathbf{E}_n(\mathbf{r}) \cdot \ort{e}_{-} \eu^{\iu \Omega_n z/c} \, \dd V, \\
		m^{\prime}_{n \mathrm{L}} &= A_n \int \limits_{\mathbb{V}_{\text{MM}}}  \left[\varepsilon(
		\vb{r})  - 1\right]\mathbf{E}_n(\mathbf{r}) \cdot \ort{e}_{+} \eu^{\iu \Omega_n z/c} \, \dd V,
	\end{split}
	\label{eq:m_prime}
\end{align}
where $\ort{e}_{\pm} = \left( \ort{x} \pm \iu \ort{y} \right) / \sqrt{2}$ are the complex unit vectors of circular polarizations, $\mathbb{V}_{\text{MM}}$ is the membrane volume, $\Omega_n =  \omega_n - \iu \gamma_n$ are  the complex eigenfrequencies, and  $\varepsilon(\vb{r})$ is the membrane material permittivity. 
We assume $\exp(-\iu \Omega_n t)$ time dependence convention, so $\gamma_n$ must be positive for passive materials. 
The factors $A_n$ depend on the modes normalization.
Indeed the electric field distributions $\mathbf{E}_n(\textbf{r})$ of the eigenmodes, as solutions of source-free equations, are naturally defined up to an arbitrary factor. To normalize them, one can define a scalar product of the modes  $(\textbf{E}_n\cdot\textbf{E}_m)$ involving numerical integration of the fields over a volume including the metasurface and over surfaces enclosing this volume \cite{Muljarov2011E,Muljarov2018OL, Lalanne2018,koshelev2022PhDthesis,Gorkunov2025AOM}. 
Then, for the modes normalized as $(\textbf{E}_n\cdot\textbf{E}_m)=\delta_{nm}$, the factors take a simple form $A_n=\iu \Omega_n \sqrt{\varepsilon_0/(2\mathbb{A}c)}$, where $\mathbb{A}$ is the area of metasurface unit cell and $c$ is the speed of light in vacuum \cite{Gorkunov2025AOM}. 

Although, very generally, nonzero $\mathrm{CD}_{\mathrm{co}}$ requires nonzero $\mathrm{CD}_{\text{mode}, n}$ for at least one mode, the particular relations between these quantities can be different, e.g. as discussed in the Supplemental Materials of Ref.~\cite{Toftul2024PRL}.
We adopt the circular dichroism  $\mathrm{CD}_{\text{mode}, n}$ and its quality factor  $Q_n =  \omega_n / (2 \gamma_n)$ as the main well defined eigenmode characteristics.

Important eigenmode properties are determined by the symmetry of MM. Thus, it is known that the rotation axis of the order 3 and higher implies that the modes of a reciprocal structure coupled to free-space waves are double degenerate \cite{Hopkins2016LPR, Kondratov2016}. For the $\mathrm{C}_4$ symmetry, the degeneracy is manifested in an intuitively clear way: the modes within such pair transform into each other upon rotation. More specifically, their field distributions $\mathbf{E}_{1}(\mathbf{r})$ and $\mathbf{E}_{\bar{1}}(\mathbf{r})$ interchange upon the action of the operator $\hat{R}_{\pi/2}$ of rotation by $\pi/2$ about the $z$-axis, (see more details in Appendix~\ref{app:symmetry}):
\begin{equation}
\begin{split}
    \hat{R}_{\pi/2} {\vb{E}}_1(\vb{r}) &= {\vb{E}}_{\bar{1}}(\vb{r}), \\
    \hat{R}_{\pi/2} {\vb{E}}_{\bar{1}}(\vb{r}) &= -{\vb{E}}_1(\vb{r}).
\end{split}
\label{eq:C4modes}
\end{equation}

The full symmetry of the single layer MM is $\mathrm{C}_{4h}$ as it includes the 'horizontal' out-of-plane mirror symmetry. This has direct consequences for the  eigenstates, as the field components have to possess certain parity with respect to the $z\rightarrow -z$ reflection (see Appendix~\ref{app:symmetry}). Following the notation from Ref.~\cite{Gorkunov2025AOM}, we denote even the modes with even in-plane field components $E_x$ and $E_y$ (and odd $E_z$), and call odd the modes with odd  $E_x$ and $E_y$ (and even $E_z$). Note that this parity immediately eliminates the mode chirality, as then, according to Eqs.~\eqref{eq:m} and \eqref{eq:m_prime}
\begin{equation}
		m_{n \mathrm{R}} = p_n m^{\prime}_{n \mathrm{L}}, \quad    
        m_{n \mathrm{L}} = p_n m^{\prime}_{n \mathrm{R}},
	\label{eq:mRLparity}
\end{equation}
where the parity indicator $p_n=1$ for an even mode and $p_n=-1$ for an odd one. 
Direct substitution of \eqref{eq:mRLparity} into the mode CD \eqref{eq:CD_mode} gives zero in the both cases, as expected.

Introducing an adjacent lossy layer to form a bilayer MM \textit{induces true geometric chirality}, enabling the system to exhibit maximal \textit{optical} chirality as quantified by the co-polarized CD, Eq.~\eqref{eq:CDco}. We note that the presence of losses is crucial for this symmetry (see also Appendix~\ref{app:CDzero}). 
While the added layer shifts the eigenfrequencies of the modes, its primary effect is to mix modes of different parities, thereby generating the mode CD, Eq.~\eqref{eq:CD_mode}, which directly contributes to the CD of the MM in transmission.

\subsection{Eigenmode transformations by symmetry breaking perturbation}
\label{sec:eigenmodes}

\begin{figure*}[t]
	\centering
	\includegraphics[width=0.9\linewidth]{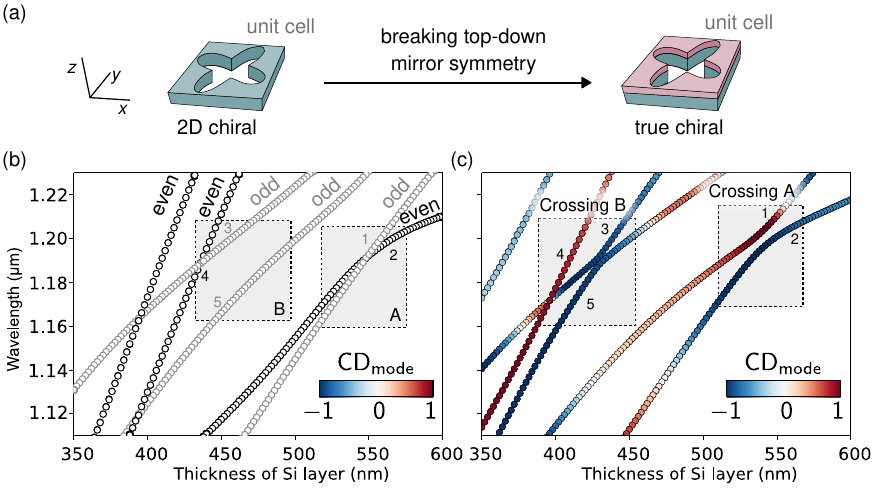}
	\caption{
		\textbf{Mode coupling via symmetry breaking. }
		(a) Variation of eigenfrequencies as a function of silicon layer thickness for a 2D chiral MM. The unit cell follows a perturbed circular profile defined in sec~\ref{sec:Design of the structure}. (b) Eigenfrequencies versus bottom Si-layer thickness for a true chiral MM, where the top layer (permittivity $\varepsilon_r = 3$) thickness of $200$~nm. 
        We note that each branch is double degenerate due to the $\mathrm{C}_4$ symmetry.}
	\label{fig:breaking_symmetry}
\end{figure*}

Employing the COMSOL eigenfrequency solver, we obtained the eigenfrequency spectrum shown in Fig.~\ref{fig:breaking_symmetry}, where each mode is double degenerate due to the MM rotation symmetry. The MM thickness is a convenient parameter to control the eigenmode spectra, and we use it to trace the evolution of modal spectral branches.

In Fig.~\ref{fig:breaking_symmetry}(b) one can see 5 spectral branches of degenerate pairs of our interest for the single layer MM. We assign them the numbers from 1 to 5 according to their parity. Thus modes 1 and 2 have opposite parity (odd and even) with respect to the mirror plane $z = 0$, and their branches intersect at Si slab thickness $580$~nm. 
Next, two other branches, labeled as 3 and 5, have same parity (odd) with respect to the mirror plane $z = 0$ and are initially well separated from each other. Their branches are intersected by another branch of the opposite (even) parity and we denote it mode 4. No mode circular dichroism is observed here as it is forbidden by the mode parity.

Next, in Fig.~\ref{fig:breaking_symmetry}(c), we show the spectral branches of a bilayer structure with two avoided crossing named as \textit{`Crossing A'}, and \textit{`Crossing B'}. Both crossings also exhibit maximum mode circular  dichroism but in different ways. 
We notice that two branches involved in Crossing A have opposite parity with respect to the $z = 0$ mirror plane and their $\mathrm{CD}_{\text{mode}}$ have opposite signs.
In turn, Crossing B involves two modes of the same parity (odd) with respect to the $z = 0$ mirror plane, and they exhibit $\mathrm{CD}_{\text{mode}}$ of the same sign.

It is important to note 
that chirality of eigenmodes ($\mathrm{CD}_{\text{mode}}$) is enhanced near avoided crossings of their spectral branches as if arising from the coupling of modes possessing either same or opposite parity. 

Understanding the origin of optical chirality requires examining how the perturbation reshapes and mixes the MM eigenmodes. In particular, we show that in the strong coupling regime, the presence of modes with \textit{opposite parity} prior to perturbation is a necessary condition.
We reveal and study two distinct cases of chiral mode mixing: 
\begin{enumerate}
    \item[(i)] Two degenerate pairs of modes of opposite parity mix [Crossing A in Fig.~\ref{fig:breaking_symmetry}(c)].
    \item[(ii)] Three degenerate pairs of modes mix, with two pairs sharing one parity and the third pair having the opposite parity [Crossing B in Fig.~\ref{fig:breaking_symmetry}(c)].
\end{enumerate}

\section{RSE theory of optical chirality }

\subsection{Chirality due to strong coupling of two spectral branches -- Crossing A}
\label{sec:twopairs}

Following the framework introduced in Ref.~\cite{Gorkunov2025AOM}, originally developed for substrate-induced mode mixing, we consider two nearly degenerate mode pairs whose strong mutual coupling dominates the response, allowing contributions from all other modes to be neglected.

To this end, the perturbation theory in therms of resonant state expansion (RSE) \cite{Muljarov2011E,Muljarov2018OL} recently generalized to account for external perturbations \cite{Almousa2023} 
allows resolving all effects of the coupling by decomposing a mode $\tilde{\textbf{E}}$ of the bilayer MM in the basis of subspace of modes ${\textbf{E}}_{m}$ of the achiral single layer MM as:
\begin{equation}
\tilde{\textbf{E}}=\sum_{m} a_m {\textbf{E}}_{m}
\label{eq:RSE}
\end{equation}
The summation goes over indices $m = 1,\bar{1},2,\bar{2}$, where the bars identify degenerate pair counterparts. The coefficients $a_m$ obey the system of equations derived directly from Maxwell's equations (see SM of Ref.~\cite{Gorkunov2025AOM}):
\begin{equation}
a_n\Omega_n=\tilde{\Omega}\sum_{m}(\delta_{nm}+V_{nm})a_m.
\label{eq:RSEsys}
\end{equation}
Here  $\Omega_{m}$ are the eigenfrequencies of modes ${\textbf{E}}_{m}$,  $\tilde{\Omega}$ is the eigenfrequency of the mode $\tilde{\textbf{E}}$ of the perturbed system, and $\delta_{nm}$ is the Kronecker delta. The elements of the  coupling parameter matrix $V_{nm}$ can be evaluated as overlap integrals of the perturbation and the corresponding modes \cite{Gorkunov2025AOM}, see also Appendix~\ref{app:Vnm} for more details. 
As all materials are isotropic dielectrics (and thus sustain Lorentz reciprocity), this matrix is symmetric, $V_{nm} = V_{mn}$. 
The initial eigenfrequencies are equal in pairs, $\Omega_1=\Omega_{\bar{1}}$ and $\Omega_2=\Omega_{\bar{2}}$, as modes $1, \bar{1}$ and $2, \bar{2}$ are degenerate and transform into each other upon rotation as in Eq.~\eqref{eq:C4modes}.

The rotation symmetry allows us to freely choose between different orthogonal linear combinations of degenerate modes. This choice can be written in a uniform way via a \textit{generalized rotation} matrix \( U(\varphi) \) as
\begin{equation}
	\begin{pmatrix}
		\tilde{\textbf{A}} \\ \tilde{\textbf{B}}
	\end{pmatrix}
	= U(\varphi) \begin{pmatrix}
		{\textbf{A}} \\ {\textbf{B}}
	\end{pmatrix}, \quad U(\varphi) = \begin{pmatrix}
		\cos \varphi & \sin \varphi \\
		-\sin \varphi & \cos \varphi
	\end{pmatrix},
	\label{eq:U}
\end{equation}
where \( \vb{A} \) and \( \vb{B} \) are any mutually orthogonal modes subjected to such transformation, 
and \( \phi \) is a \textit{generalized angle} which can  be complex. The rotation matrix satisfies \( U^{T}(\varphi) U(\varphi) = 1 \), hence preserving the norm and orthogonality during such rotation. If the initial pair of modes satisfied the property \eqref{eq:C4modes}, so will do the pair created by the transformation \eqref{eq:U}.

To describe the mode transformations induced by adding a second MM layer, we start with an arbitrary set of two pairs of orthogonal degenerate modes  ${\textbf{E}}_1$, ${\textbf{E}}_{\bar{1}}$ and ${\textbf{E}}_2$, ${\textbf{E}}_{\bar{2}}$ corresponding to the eigenfrequencies $\Omega_1$ and \( \Omega_{2} \) respectively. 
Coupling between degenerate modes is absent, $V_{1\bar{1}} = V_{2\bar{2}} = 0$, which can be shown by applying the symmetry property~\eqref{eq:C4modes} under the integral that defines $V_{nm}$ (see Appendix~\ref{app:Vnm}).
Then, for a particular structural perturbation retaining the rotation symmetry, it is possible to chose the pair of orthogonal degenerate modes ${\textbf{E}}_{2}$ and ${\textbf{E}}_{\bar{2}}$ using \eqref{eq:U} to establish $V_{1\bar{2}}=V_{\bar{1}2}=0$, i.e., so that the perturbation couples only mode $1$ with mode $2$ and, symmetrically, mode $\bar{1}$ with mode $\bar{2}$. This considerably simplifies the analysis, as it essentially reduces the problem to a pair of strongly coupled oscillators. The final available degree of freedom, namely the definition of modes ${\textbf{E}}_1$ and ${\textbf{E}}_{\bar{1}}$, will be fixed to simplify the parametrization of the coupling parameters \eqref{eq:m} and \eqref{eq:m_prime} later in this section.

The perturbed mode must also be normalized, $(\tilde{\mathbf{E}} \cdot \tilde{\mathbf{E}}) = 1$, which fixes both its amplitude and phase (see Appendix~\ref{app:gen_rot} and Refs.~\cite{Muljarov2011E,Muljarov2018OL, Lalanne2018,koshelev2022PhDthesis,Gorkunov2025AOM}). This imposes a constraint on the coefficients in Eq.~\eqref{eq:RSE}: $a_{1}^2 + a_{2}^2  = a_{\bar{1}}^2 + a_{\bar{2}}^2 = 1$ and allows writing the mixed mode in a form similar to the transformation \eqref{eq:U}:
\begin{equation}
\tilde{\textbf{E}}=\cos\phi\ \textbf{E}_{1}+\sin\phi\ {\bf E}_{2},
\label{eq:E12}
\end{equation}
where the angle $\phi$ (also generally complex) characterizes the degree of mode mixing and we call it a \textit{mixin angle}.

Accordingly, the eigenvalue problem \eqref{eq:RSEsys} is reduced to the system:
\begin{equation}
    \begin{pmatrix}
        \tilde\Omega-\tilde\Omega_1 & \dfrac{\tilde{\Omega} V_{21}}{1+V_{11}} \\ \ \\
        \dfrac{\tilde{\Omega} V_{21}}{1+V_{22}} & \tilde\Omega-\tilde\Omega_2
    \end{pmatrix} \begin{pmatrix}
        \cos \phi \\ \ \\ \ \\   \sin \phi
    \end{pmatrix} = 0
    \label{eq:RSE1}
\end{equation}
yielding a pair of mixed mode eigenfrequencies:
\begin{equation}
    \tilde{\Omega}_{\pm} = \frac{\tilde\Omega_1 +\tilde\Omega_2	\pm \sqrt{(\tilde\Omega_1 -\tilde\Omega_2)^2+4 u^2\tilde\Omega_1\tilde\Omega_2}}{2(1-u^2)}.
	\label{eq:freqsplit}
\end{equation}
Here the introduced frequencies $\tilde\Omega_{1}=\Omega_{1}/(1+V_{11})$ and $\tilde\Omega_{2}=\Omega_{2}/(1+V_{22})$ describe the perturbation-induced shift of the mode eigenfrequencies regardless of the mode coupling, while the parameter $u=V_{12}/\sqrt{(1+V_{11})(1+V_{22})}$ characterizes the coupling contribution.

The corresponding mixing angles are defined as
\begin{equation}
    \phi_\pm=\atan \frac{(\tilde\Omega_1-\tilde\Omega_\pm)(1+V_{11})}{\tilde\Omega_\pm V_{12}} 
    \label{eq:mixangle}
\end{equation}
and it is straightforward to obtain that 
\begin{equation}
    \tan\phi_+\tan\phi_-=-\frac{\Omega_1}{\Omega_2},
    \label{eq:phiphi}
\end{equation}
i.e., as the considered initial modes possess very close eigenfrequencies, $|\Omega_2-\Omega_1| \ll \Omega_{1,2}$, the angles $\phi_+$ and $\phi_-$ approximately differ by $\pi/2$. Therefore, we can express the pair of mixed modes $\tilde{\textbf{E}}_{\pm}$ corresponding to the frequencies $\tilde{\Omega}_{\pm}$ as
\begin{equation}
\begin{pmatrix}
    \tilde{\textbf{E}}_{+} \\ \tilde{\textbf{E}}_{-}
\end{pmatrix}
= U(\phi_+) \begin{pmatrix}
    {\textbf{E}}_{1} \\ {\textbf{E}}_{2}
\end{pmatrix},
\label{eq:tE12}
\end{equation}
which preserves the norm and orthogonality of the modes, see also Appendix~\ref{app:gen_rot}. 
Note that as the angles $\phi_{+}$ or $\phi_{-}$  differ by $\pi/2$, one can symmetrically express 
\begin{equation}
\begin{pmatrix}
    \tilde{\textbf{E}}_{-} \\ \tilde{\textbf{E}}_{+}
\end{pmatrix}
= U(\phi_-) \begin{pmatrix}
    {\textbf{E}}_{1} \\ {\textbf{E}}_{2}
\end{pmatrix}.
\label{eq:tE12minus}
\end{equation}

As the adjacent layer breaks the out-of-plane symmetry while preserving the $C_4$ symmetry, the eigenmodes remain degenerate in pairs and are related within these pairs by the property~\eqref{eq:C4modes}. The corresponding degenerate counterparts of the mixed modes $\tilde{\textbf{E}}_{\pm}$ can be expressed by similar generalized rotations:
\begin{equation}
\begin{pmatrix}
    \tilde{\textbf{E}}_{\bar+} \\ \tilde{\textbf{E}}_{\bar-}
\end{pmatrix}
= U(\phi_+) \begin{pmatrix}
    {\textbf{E}}_{\bar1} \\ {\textbf{E}}_{\bar2}
\end{pmatrix}, \ {\rm or}
\begin{pmatrix}
    \tilde{\textbf{E}}_{\bar-} \\ \tilde{\textbf{E}}_{\bar+}
\end{pmatrix}
= U(\phi_-) \begin{pmatrix}
    {\textbf{E}}_{\bar1} \\ {\textbf{E}}_{\bar2}
\end{pmatrix}.
\label{eq:bartE12}
\end{equation}
Altogether, we conclude that the perturbation effectively mixes the modes in the form of generalized rotation \eqref{eq:U} in their subspaces.

For absent mode mixing, $u=0$, according to Eq.~\eqref{eq:freqsplit}, the eigenfrequency $\tilde{\Omega}_+$  turns into the largest of $\tilde{\Omega}_1$ and $\tilde{\Omega}_3$, while $\tilde{\Omega}_-$ equals to the smaller of them. Importantly, the real parts of the eigenfrequencies are free to intersect in the parameter space, and the corresponding values of $\phi_{+}$ and $\phi_{-}$ being equal to either $0$ or $\pi/2$ ensure that the field profiles of the modes remain unaffected by the perturbation. 

For nonzero coupling $V_{12}$, even if for a certain set of structure parameters the real parts of the frequencies $\tilde{\Omega}_{1,2} = \tilde{\omega}_{1,2} - \iu \tilde{\gamma}_{1,2}$ become equal, $\tilde\omega_1=\tilde\omega_2$, the intersection of the two spectral branches given by Eq.~\eqref{eq:freqsplit} remains forbidden. 
Instead, mode anti-crossing is to be observed, 
if the corresponding resonance widths allow for such spectral resolution, i.e., when   
\begin{equation}
	|u|> \frac{1}{\operatorname{max}(\tilde\omega_{1}, \tilde\omega_{2})}\sqrt{\frac{\tilde\gamma_1^2+\tilde\gamma_2^2}{2}}.
	\label{eq:strong coupling1}
\end{equation}
Note that, at the same time, as it follows from Eq.~\eqref{eq:mixangle}, a considerable mode mixing with $\phi_{\pm} \simeq \pi/4$ accompanied by a characteristic Rabi splitting of the modes $\tilde\Omega_\pm$ occurs already when 
\begin{equation}
    	|u|> \frac{1}{\operatorname{max}(\tilde\omega_{1}, \tilde\omega_{2})}{|\tilde\gamma_1-\tilde\gamma_2|}, 
	\label{eq:strong coupling}
\end{equation}
which is the criteria of strong coupling~\cite{SAVONA1995733:Strongcoupling} for the case considered here. 

It is also possible to obtain rather compact expressions for the coupling parameters of the mixed modes assuming that the initial modes of the single layer MM have linearly polarized far field asymptotics. 
For this purpose we utilize the remaining degree of freedom to choose a convenient mixing angle between first degenerate pair of $1,\bar{1}$ modes using transformation \eqref{eq:U} such that $m_{1,R} = m_{1,L}$.
A final assumption is that mode $2$ is linearly polarized in the far field with the angle $\psi$ relative to the mode $1$.
More details are in Appendix~\ref{app:mRL}. 
Thus, we reduce 8 complex coupling parameters of all initial modes to only two complex amplitudes $M_1$, $M_2$ and an angle $\psi$:
\begin{align}
\begin{split}
    m_{1,R} = m_{1,L}= p_1m'_{1,R} = p_1m'_{1,L} & = M_1, \\
    m_{2,R} = p_2 m'_{2,L} &= M_2 \eu^{\iu \psi}, \\ 
    m_{2,L} = p_2 m'_{2,R} & = M_2 \eu^{-\iu\psi}. 
\end{split}
\label{eq:M1M2}
\end{align} 
\begin{figure}[t]
    \centering
    \includegraphics[width = 0.9\linewidth]{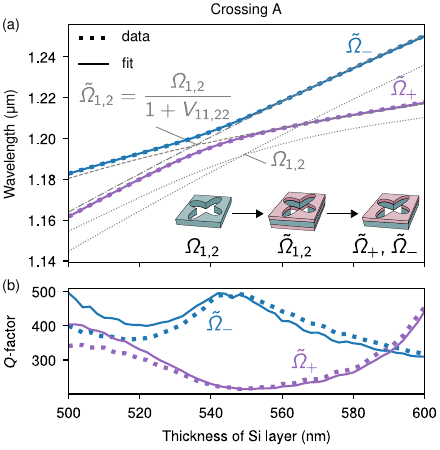}
    \caption{\textbf{Fitting of crossing A.} (a) Real part of the wavelength. Dotted lines indicate the modes (\(\Omega_{1(2)}\)) of the single Si layer. Dashed lines represent the modes  (\(\tilde\Omega_{1(2)}\)) for a symmetric perturbation, where an identical second layer is added on both sides with \(\Delta\epsilon =1\)),  Coloured solid lines show the fitted modes (\(\Omega_{\pm}\)) using Eq.~(\ref{eq:freqsplit}), while the dotted lines correspond to numerically calculated modes for the asymmetric perturbation with \(\Delta\epsilon =2\)), (b) Numerically computed Q-factors (dots) and Q-factors obtained from fitting (solid lines).}
    \label{fig:crossingAfit}
\end{figure}
For the mode mixed as described by Eq.~\eqref{eq:tE12}, using \eqref{eq:M1M2} we obtain the coupling parameters as:
\begin{align}
\begin{split}
    \begin{pmatrix}
        \tilde{m}_{+R} \\ \tilde{m}_{-R}
    \end{pmatrix} &= U(\phi_+)
    \begin{pmatrix}
        M_1 \\ M_2 \eu^{\iu \psi}
    \end{pmatrix}, \\ 
    \begin{pmatrix}
        \tilde{m}_{+L} \\ \tilde{m}_{-L}
    \end{pmatrix} &= U(\phi_+)
    \begin{pmatrix}
        M_{1} \\ M_2 \eu^{-\iu \psi}
    \end{pmatrix}, \\
    \begin{pmatrix}
        \tilde{m}^{\prime}_{+R} \\ \tilde{m}^{\prime}_{-R}
    \end{pmatrix} &= U(\phi_+)
    \begin{pmatrix}
        p_1 M_1 \\ p_2 M_2 \eu^{-\iu \psi}
    \end{pmatrix}, \\
    \begin{pmatrix}
        \tilde{m}_{+L}^{\prime} \\ \tilde{m}_{-L}^{\prime}
    \end{pmatrix} &= U(\phi_+)
    \begin{pmatrix}
        p_1 M_1 \\ p_2 M_2 \eu^{\iu \psi}
    \end{pmatrix}.
\end{split}
\label{eq:mtilde12}
\end{align}
Now it is possible to estimate potential degree of mode CD. We find that
\begin{align}
\begin{split}
    |\tilde{m}_{+R}|^2 &= C + \Re \left(M_1M_2^*\eu^{-\iu\psi}\right) \sin 2 \phi_+, \\
    |\tilde{m}_{+L}|^2 &= C + \Re \left(M_1M_2^*\eu^{\iu\psi}\right) \sin 2 \phi_+, \\ 
    |\tilde{m}^{\prime}_{+R}|^2 &= C + p_1 p_2 \Re \left(M_1M_2^*\eu^{-\iu\psi} \right) \sin 2 \phi_+, \\
    |\tilde{m}^{\prime}_{+L}|^2 &= C + p_1 p_2 \Re  \left(M_1M_2^*\eu^{\iu\psi}\right) \sin 2 \phi_+,
\end{split}
\end{align}
where $C = |M_1|^2\cos^2\phi_++|M_2|^2\sin^2\phi_+$, and we have assumed that mixing angle $\phi$ has a negligible imaginary part. 
This allows obtaining the nominator of the mode CD \eqref{eq:CD_mode} as
\begin{multline}
|\tilde{m}_{+\mathrm{R}} \tilde{m}^{\prime}_{+ \mathrm{R}}|^2 - |\tilde{m}_{+\mathrm{L}} \tilde{m}^{\prime}_{+\mathrm{L}}|^2=(1-p_1p_2)\sin\psi\\
\times{\rm Im}\left[\sin2\phi_+M_1M_2
\left({M_2^{*}}^2\sin^2\sin2\phi_+\right.\right.\\
\left.\left.
-{M_1^{*}}^2\cos^2\sin2\phi_+\right)\right],
\label{eq:CD_mode_1}
\end{multline}
hence
\begin{equation}
    \mathrm{CD}_{\text{mode},+} \propto (1-p_1p_2).
\label{eq:CD_mode_parity}
\end{equation}
One can conclude that in this case nonzero CD can occur only if the initial degenerate pairs have \textit{opposite parity} with $p_1p_2=-1$. 

According to its definition \eqref{eq:CD_mode}, the mode CD achieves its extreme $+1$ ($-1$) value, if one of the coupling parameters $m_{nL}$ or $m'_{nL}$ ($m_{nR}$ or $m'_{nR}$) turns to zero while all such for the opposite circular polarization remain finite. Taking as example the parameters of the mixed mode $\tilde{\textbf{E}}_{+}$ given by Eqs.~\eqref{eq:mtilde12}, it is easy to see that its $\mathrm{CD}_{\text{mode},+}=1$ when
\begin{equation}
    \eu^{\pm\iu\psi}=\mp\frac{M_1}{M_2}\cot{\phi_+}.
\end{equation}
This is incompatible with the other mixed mode $\tilde{\textbf{E}}_{-}$ also having $\mathrm{CD}_{\text{mode},-}=1$, but it can occur together with  
\begin{equation}
    \eu^{\pm\iu\psi}=\mp\frac{M_1}{M_2}\tan{\phi_+},
\end{equation}
which is the condition of this mixed mode having the opposite optical chirality with $\mathrm{CD}_{\text{mode},-}=-1$ when the mixing angle $\phi_+=\pm\pi/4$.

The fact that strongly mixed modes here are bound to exhibit optical chirality of opposite signs can be traced from the form of coupling parameters given by Eqs.~\eqref{eq:mtilde12}.
For example, considering for definiteness the case of strong mixing with $\phi_+=\pi/4$ of modes of opposite parity with $p_1=-1$ and $p_2=1$, these equations yield $\tilde{m}_{+R}=\tilde{m}'_{-L}$, $\tilde{m}'_{+R}=\tilde{m}_{-L}$, etc., i.e., the mixed modes $\tilde{\textbf{E}}_{+}$ and $\tilde{\textbf{E}}_{-}$ are equally coupled to circularly polarized waves of opposite helicity incident on the opposite MM sides. This generally leads to $\mathrm{CD}_{\text{mode},1}=-\mathrm{CD}_{\text{mode},2}$.

With Eq.~\eqref{eq:CD_mode_parity} being  one of the the central results of this paper, we conclude that the coupling of two  pairs of degenerate modes gives rise to optical chirality only when their initial parities are opposite. Then two pairs of degenerate mixed modes can reach maximum mode CD of opposite signs.

In order to illustrate how the above scenario of mode coupling is realized in the simulated Crossing A, we perform fitting with Eq.~\eqref{eq:freqsplit} the numerically obtained spectral branches of modes. To minimize the fitting ambiguity, we use the eigenfrequencies $\Omega_{1,2}$ obtained for the unperturbed single layer MM, and also use the results of simulations of MM in symmetrically perturbed environment of  \textit{the same strength} such that self interaction coefficients $V_{11,22}$ for symmetric and antisymmetric are approximately equal (see Appendix~\ref{app:Vnm}). Thus, such symmetric perturbation gives an approximated values of $\tilde{\Omega}_{1,2}$. The latter allow directly resolving the self-action of modes determined by the perturbation matrix elements $V_{11,22} = \Omega_{1,2} / \tilde{\Omega}_{1,2} - 1$. Then there remains only one element $V_{12}$ to be determined from the fitting, and, as is shown in Fig.~\ref{fig:crossingAfit}, it is possible to reconstruct rather accurately both real and imaginary parts of $\tilde{\Omega}_{\pm}$ assuming reasonable values of $V_{12}$ for the considered range of MM thickness. For technical details of the fitting and the corresponding values of $V_{nm}$ resolved as functions of the Si layer thickness see Appendix~\ref{app:fitting}.

\subsection{Chirality due to coupling of three spectral branches -- Crossing B}
\label{sec:threepairs}
Consider now a peculiar case of Crossing B  observed by simulations and shown in Fig.~\ref{fig:breaking_symmetry}(c) where two pairs of degenerate modes of the same parity (odd modes 3,$\bar3$ and 5,$\bar5$) are strongly coupled and possess large modal CD of the same sign. According to Eq.~\eqref{eq:CD_mode_parity}, optical chirality cannot arise due to direct coupling of modes of the same parity with $p_3=p_5$. Therefore, to explain the observations, one has to include into consideration a third pair of degenerate modes of the opposite parity (even modes 4,$\bar4$) which is indeed observed by simulations in the vicinity of Crossing B.

Useful details of mode coupling can be extracted from comparing the mode branch transformations occurring when symmetric and asymmetric perturbations are applied to the same MM, as in described in Appendix~\ref{app:Vnm}. In particular, the simulations reveal the lack of direct coupling of modes (3,$\bar3$) and (5,$\bar5$) of the same parity with $V_{35} \simeq V_{3\bar5} \simeq  V_{\bar35} \simeq  V_{\bar3\bar5} \simeq 0$, which would otherwise manifest itself for the symmetric perturbation. 
For asymmetric perturbation, these branches form an anti-crossing with a relatively weak Rabi splitting. We demonstrate below, that this can be explained by weak indirect coupling of these modes via another pair of modes (4,$\bar4$) having different parity.

Extending the RSE approach to account for more modes in the sums in Eqs.~\eqref{eq:RSE} and \eqref{eq:RSEsys}, one can write an  analog of Eq.~\eqref{eq:E12} as:
\begin{equation}
\tilde{\textbf{E}}=\cos\theta\cos\phi\ \textbf{E}_{3}+\cos\theta\sin\phi\ {\bf E}_{5}\\
+\sin\theta {\bf E}_{4},
\label{eq:E123}
\end{equation}
where $\phi$ and $\theta$ are the new complex mixing angles.
As discussed in Appendix~\ref{app:3pairs}, the rotation symmetry of the MM and the perturbation allows writing mixed modes in such form, provided that the states 3 and 5 are specially chosen to be uncoupled from mode $\bar4$. 
Besides, we have used the fact the initial modes 3 and 5 are also orthogonal, $(\mathbf{E}_3 \cdot \mathbf{E}_5) = 0$ due the properties of the resonant states and corresponding definition of the scalar product~\cite{Lalanne2018LPR,Gorkunov2025AOM}. This simplifies constrain on the expansion coefficients in Eq.~\eqref{eq:RSE}: $a_3^2 + a_4^2 + a_5^2 = 1$.
Thus, introducing a pair of mixing angles $\phi$ and $\theta$ now preserves the mode \eqref{eq:E123} normalization.

In the absence of direct coupling of modes (3,$\bar3$) and (5,$\bar5$), an extended analog of the eigenvalue problem \eqref{eq:RSEsys} 
for the expansion \eqref{eq:E123} yields a system of equations: 
\begin{align}
\left[\Omega_3-\tilde\Omega(1+V_{33})\right]\cos\phi
&=\tilde\Omega V_{34}\tan\theta,
\label{eq:3RSE1}\\
\left[\Omega_5-\tilde\Omega(1+V_{55})\right]\sin\phi
&=\tilde\Omega V_{45}\tan\theta,
\label{eq:3RSE3}\\        
\left[\Omega_4-\tilde\Omega(1+V_{44})\right]\tan\theta &=\tilde\Omega( V_{34}\cos\phi+ V_{45}\sin\phi),
\label{eq:3RSE5}        
\end{align}
Substituting $\tan\theta$ from \eqref{eq:3RSE5} into Eqs.~\eqref{eq:3RSE1}--\eqref{eq:3RSE3} yields an analog of Eq.~\eqref{eq:RSE1}:
\begin{equation}
    \begin{pmatrix}
        \tilde\Omega-\tilde\Omega_3+\dfrac{v_{3}\tilde\Omega^2}{\tilde\Omega_4-\tilde\Omega} & \dfrac{w_{3}\tilde\Omega^2}{\tilde\Omega_4-\tilde\Omega} \\ \ \\ 
        \dfrac{w_{5}\tilde\Omega^2}{\tilde\Omega_4-\tilde\Omega} & \tilde\Omega-\tilde\Omega_5+\dfrac{v_{5}\tilde\Omega^2}{\tilde\Omega_4-\tilde\Omega}
    \end{pmatrix} \begin{pmatrix}
        \cos \phi \\ \ \\  \sin \phi
    \end{pmatrix} = 0,
    \label{eq:6RSE1}
\end{equation}
where the shifted eigenfrequencies, $\tilde\Omega_n = \Omega_n / (1 + V_{nn})$ with $n=3,4,5$,  are introduced.
The coupling parameters are
\begin{align}
\begin{split}
    v_{3}&=\frac{V_{34}^2}{(1+V_{33})(1+V_{44})}, \ 
    v_{5}=\frac{V_{45}^2}{(1+V_{55})(1+V_{44})},\\
    w_{3}&=\frac{V_{34}V_{45}}{(1+V_{33})(1+V_{44})}, \ 
    w_{5}=\frac{V_{34}V_{45}}{(1+V_{55})(1+V_{44})}.
\end{split}
\end{align}
Considering the frequency range around the spectral branch crossing of modes 3 and 5, i.e., assuming that $\tilde\Omega\simeq\tilde\Omega_3\simeq\tilde\Omega_5$ one can further simplify Eqs.~\eqref{eq:6RSE1} to 
\begin{equation}
    \begin{pmatrix}
        \tilde\Omega-\bar{\bar\Omega}_3 & \tilde\Omega u_3 \\
        \tilde\Omega u_5& \tilde\Omega-\bar{\bar\Omega}_5
    \end{pmatrix} \begin{pmatrix}
        \cos \phi \\ \sin \phi
    \end{pmatrix} = 0
    \label{eq:eff6RSE1}
\end{equation}
where 
\begin{equation}
\bar{\bar\Omega}_{3(5)}=\tilde\Omega_{3(5)}-\frac{v_{3(5)}\tilde\Omega_{3(5)}^2}{\tilde\Omega_4-\tilde\Omega_{3(5)}}
\label{eq:barOmega}
\end{equation}
are the frequencies of pairs of modes ($3,\bar3$) and ($5,\bar5$) shifted by the coupling to mode 4, and the effective coupling parameters are:
\begin{equation}
u_{3(5)}=\frac{w_{3(5)}\tilde\Omega_{3(5)}^2}{\tilde\Omega_4-\tilde\Omega_{3(5)}}.
\label{eq:u13}
\end{equation}

Comparing the system of Eqs.~\eqref{eq:eff6RSE1} with analogous Eqs.~\eqref{eq:RSE1} written for directly coupled modes, one can see that the indirect interaction through the third pair produces additional frequency shifts and also the coupling which are both generally weak being proportional to higher powers of the interaction matrix elements $V_{nm}$. On the other hand, when the frequency $\tilde\Omega_4$ approaches the crossing point of branches $\tilde\Omega_3$ and $\tilde\Omega_5$, the effects of indirect coupling are enhanced as the denominators in Eqs.~\eqref{eq:barOmega}--\eqref{eq:u13} decrease.  

Solving the system~\eqref{eq:eff6RSE1}, one obtains the anti-crossing of branches of modes ($3,\bar3$) and ($5,\bar5$) described by an analog of Eq.~\eqref{eq:freqsplit}:
\begin{equation}
    \tilde{\Omega}_{\pm} = \frac{\bar{\bar\Omega}_3 +\bar{\bar\Omega}_5	\pm \sqrt{(\bar{\bar\Omega}_3 -\bar{\bar\Omega}_5)^2+4 u_3u_5\bar{\bar\Omega}_3\bar{\bar\Omega}_5}}{2(1-u_3u_5)}.
	\label{eq:freqsplit2}
\end{equation}
Similarly to the case of directly coupled modes, here also the corresponding mixing angles are
\begin{equation}
    \phi_\pm=\atan \frac{\bar{\bar{\Omega}}_3-\tilde\Omega_\pm}{\tilde\Omega_\pm u_3} 
    \label{eq:mixangle2}
\end{equation}
and, similarly to Eq.~\eqref{eq:phiphi}, here for the mixing angles corresponding to different branches one can derive that
\begin{equation}
    \tan\phi_+\tan\phi_-=-\frac{u_5\bar{\bar{\Omega}}_3}{u_3\bar{\bar{\Omega}}_5},
    \label{eq:phiphi2}
\end{equation}
i.e., as the frequencies $\bar{\bar{\Omega}}_3$ and $\bar{\bar{\Omega}}_5$ are close, the angles $\phi_+$ and $\phi_-$ approximately differ by $\pi/2$, and the indirect coupling of pairs of modes ($3\bar3$) and ($5\bar5$) effectively causes their generalized rotation as in Eq.~(\ref{eq:tE12}).
\begin{figure}
    \centering
    \includegraphics[width=0.9\linewidth]{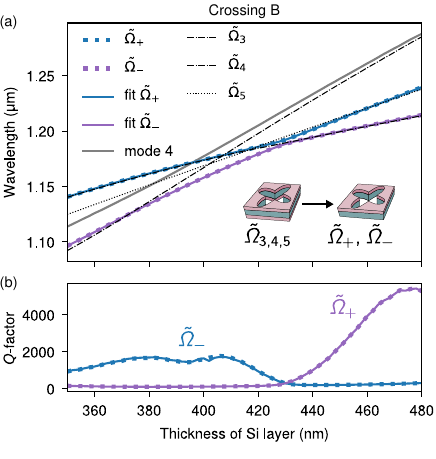}
    \caption{\textbf{Fitting of crossing B.} (a) Real part of the wavelength. Dashed lines indicate the modes (\(\tilde{\Omega}_{1,2,3}\)) of the symmetrically perturbed Si layer (same as in Fig.~\ref{fig:crossingAfit}).  Coloured solid lines show the fitted modes (\(\Omega_{\pm}\)) using Eq.~(\ref{eq:freqsplit2}). (b) Numerically computed Q-factors (dots) and Q-factors obtained from fitting (solid lines).}
    \label{fig:crossingBfit}
\end{figure}
Mixed eigenmodes given by Eq.~\eqref{eq:E123} then can be expressed as: 
\begin{multline}
\tilde{\textbf{E}}_\pm=\cos\theta_\pm \\
\times\left[\cos\phi_\pm\left({\bf E}_{3}+q_3{\bf E}_{4} \right)+\sin\phi_\pm\left({\bf E}_{5}+q_5{\bf E}_{4} \right)\right],\label{eq:E135fin}
\end{multline}
where the mixing constants were defined as
\begin{equation}
    q_{3(5)}=\frac{\tilde\Omega_{3(5)}}{\tilde\Omega_4-\tilde\Omega_{3(5)}}\frac{V_{3(5)4}}{1+V_{44}},
\end{equation}
Eq.~\eqref{eq:3RSE5} was used to obtain $\tan\theta_\pm$, and the corresponding angles $\theta_\pm$ can be found from there if necessary too. 

One can see that the modes $\tilde{\textbf{E}}_\pm$ corresponding to both anticrossing spectral branches consist of linear superpositions of modes 
\begin{equation}
\tilde{\bf E}_{3(5)}={\bf E}_{3(5)}+q_{3(5)}{\bf E}_{4}    
\end{equation}
which, similarly to the mixture of modes described by Eq.~\eqref{eq:E12} for Crossing A, are built of modes of different parity. In the same way, their far-field chirality can be resolved from analyzing superpositions of the constituent far field asymptotics: 
\begin{align}
\begin{split}
    \tilde{m}_{3(5),R} &=  M_{3(5)}+q_{3(5)}M_4\eu^{\iu\psi} , \\
    \tilde{m}_{3(5),L} &=  M_{3(5)}+q_{3(5)}M_4\eu^{-\iu \psi} , \\
    \tilde{m}^{\prime}_{3(5),R} &=  p_3M_{3(5)}+p_4q_{3(5)}M_4\eu^{-\iu \psi} , \\
    \tilde{m}^{\prime}_{3(5),L} &=  p_3M_{3(5)}+p_4q_{3(5)}M_4\eu^{\iu\psi} , 
\end{split}
\label{eq:tm13R}
\end{align}
where, similarly to the case described in Appendix~\ref{app:mRL}, the coupling parameters of the initial modes are expressed by the complex amplitudes $M_{3(5)}$ and $M_4$ and an angle $\Psi$. 

Qualitative differences of both types of crossings are evident from comparing the latter Eqs.~\eqref{eq:tm13R} with the former Eqs.~\eqref{eq:mtilde12}. The coupling of pairs of modes ($1,\bar1$) and ($2,\bar2$) for Crossing A were bound to lead to opposite signs of their ${\rm CD}_{\rm mode}$. For Crossing B, the situation is much more flexible, as the sings of optical chirality of modes $\tilde{\bf E}_{3(5)}$, in general, are not subjected to any bounds. In the case when they are of the same sign, their linear superpositions in the actual mixed modes \eqref{eq:E135fin} constructively interfere giving a rise to a pair of strongly chiral optical resonances with the same sign of CD. 

Another interesting feature of the indirect coupling is that the mixing constants $q_{3(5)}$ accompanying mode 4 in the coupling parameters \eqref{eq:tm13R} are determined by the perturbation and, therefore, should be relatively small. This smallness, however, is naturally compensated by much stronger far field asymptotics of mode 4 which quality factor is by at least  an order of magnitude smaller than that of modes 3 and 5. Accordingly, all modes contribute to Eqs.~\eqref{eq:tm13R} comparably strongly and this enables strong CD of the mixed modes.

To illustrate how the above specific scenario of coupling of three spectral branches is realized within Crossing B, we perform again the  fitting of the numerically obtained $\tilde{\Omega}_{\pm}$ with the model Eq.~\eqref{eq:freqsplit2}. Compared to Crossing A, more  elements of the perturbation matrix are to be determined here. We use again the eigenfrequencies of unperturbed single-layer MM to set $\Omega_{3,4,5}$ and use the simulations for the symmetric perturbation to determine the elements $V_{33}$, $V_{44}, V_{55}$. Then two elements $V_{34}$ and $V_{45}$ remain to be obtained by the fitting, and, as one can see in Fig.~\ref{fig:crossingBfit}, one can  reconstruct real and imaginary parts of $\tilde{\Omega}_{\pm}$ with a very good accuracy. Further technical details and the corresponding values of $V_{nm}$ are provided in Appendix~\ref{app:fitting}.

\section{Loss engineering to maximize optical chirality}

\begin{figure}
    \centering
    \includegraphics[width=0.90\linewidth]{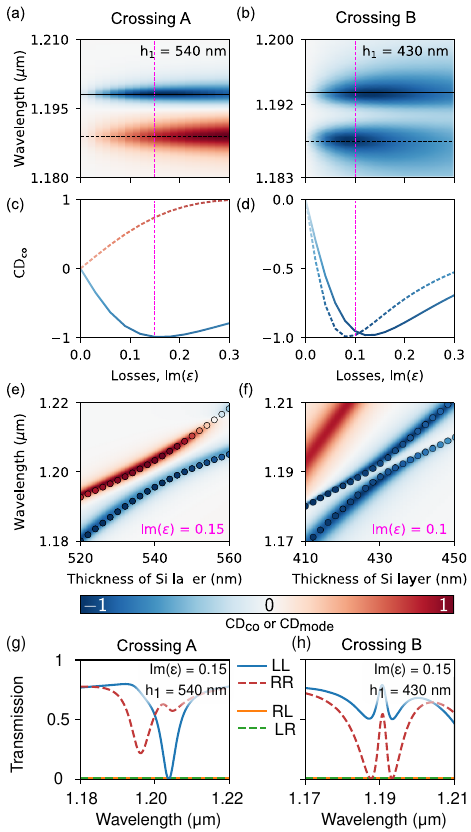}
    \caption{\textbf{Loss-induced circular dichroism.} (a, b) Co-polarized circular dichroism ($\mathrm{CD}_{\text{co}}$) as a function of the imaginary part of the permittivity introduced in the second layer ($\varepsilon_2 = 3 + \iu \Im\varepsilon$) for a structure with first layer thickness $h_1 = 540$~nm [panel (a)] and $h_1 = 430$~nm [panel (b)]. The second layer thickness is $h_2 = 200$~nm for both cases. $\mathrm{CD}_{\text{co}}$ values at two resonance wavelengths are marked by horizontal lines in (a,b) panels. (c, d) Each resonance reaches its maximum $\mathrm{CD}_{\text{co}}$ at a different value of $\Im \varepsilon$, indicating the occurrence of critical coupling. 
    (e, f) $\mathrm{CD}_{\text{co}}$ (color map) and $\mathrm{CD}_{\text{mode}}$ (colored circles) at the mode anti-crossings between two modes of the same parity for $z=0$ plane mirror symmetry, observed at $\Im \varepsilon = 0.15$ [panel (e)] and of the different parity observed at $\Im\varepsilon = 0.1$ [panel (f)]. The transmission spectra for Crossing A (g) and Crossing B (h) for optimal levels of dissipation loss.}
    \label{fig:optimal_losses}
\end{figure}

Since for the $\mathrm{C}_4$ symmetric reciprocal metasurface the reflection is helicity independent and there is no cross-polarization conversion in transmission (Appendix~\ref{app:CDzero}), we can write \textit{absorption} coefficients as
\begin{equation}
    A_{\text{R}} = 1 - R - T_{\text{R}}, \qquad 
    A_{\text{L}} = 1 - R - T_{\text{L}},
    \label{eq:A}
\end{equation}
where $R = \abs{r_{\text{LR}}}^2 = \abs{r_{\text{RL}}}^2$ is the reflection coefficient, $T_{\text{R}} = \abs{t_{\text{RR}}}^2$ and $T_{\text{L}} = \abs{t_{\text{LL}}}^2$ are the transmission coefficients for RCP and LCP light.
Substituting Eq.~\eqref{eq:A} into Eq.~\eqref{eq:CDco} leads to the following form of the co-polarized circular dichroism:
\begin{equation}
    \mathrm{CD}_{\mathrm{co}} =  \frac{A_{\text{R}} - A_{\text{L}}}{2 (1-R) - (A_{\text{R}} + A_{\text{L}})}.
    \label{eq:CDcoABS}
\end{equation}
Using chiral coupled mode theory, one can relate absorption coefficients in Eq.~\eqref{eq:A} with mode coupling coefficients in Eqs.~\eqref{eq:m} in the vicinity of the resonance as~\cite{Gorkunov2020PRL}
\begin{equation}
    \begin{split}
        A_{\text{R}} \simeq  \frac{2\gamma_{\text{abs}} |m_R|^2}{(\omega - \omega_0)^2 + \gamma_0^2}, \quad 
        A_{\text{L}} \simeq  \frac{2\gamma_{\text{abs}} |m_L|^2}{(\omega - \omega_0)^2 + \gamma_0^2},
    \end{split}
    \label{eq:AR_AL}
\end{equation}
where resonance is characterised by the complex eigen frequency $\omega_0 - \iu \gamma_0$, where total decay rate $\gamma_0$ can be decomposed into radiation and absorption losses: $\gamma_0 \simeq \gamma_{\text{rad}} + \gamma_{\text{abs}}$. We note that this decomposition is feasible only for the high-$Q$ modes, where $Q = \omega_0/(2\gamma_0)$~\cite{Qfactor_1,Qfactor2}.

At a close-to-maximum chiral resonance, radiation of one helicity in all channels is significantly suppressed, e.g., $m_{\text{LL}} \simeq m^{\prime}_{\text{LL}} \simeq 0$, which, in particular  leads to unit mode circular dichroism, $\mathrm{CD}_{\text{mode}} = 1$. Substituting \eqref{eq:AR_AL} into \eqref{eq:CDco} at the resonance frequency $\omega \simeq \omega_0$, assuming that LCP channels are suppressed, we arrive at
\begin{equation}
\begin{split}
    \mathrm{CD}_{\mathrm{co}} &\simeq \frac{ \gamma_{\text{abs}}|m_{\text{R}}|^2}{(1 - R) (\gamma_{\text{rad}} + \gamma_{\text{abs}})^2 - |m_{\text{R}}|^2 \gamma_{\text{abs}}} \\
    &\simeq \frac{ \gamma_{\text{abs}}|m^{\prime}_{\text{R}}|^2}{(1 - R^{\prime}) (\gamma_{\text{rad}} + \gamma_{\text{abs}})^2 - |m^{\prime}_{\text{R}}|^2 \gamma_{\text{abs}}}.
\end{split}
\end{equation}
Condition $\partial_{\gamma_{\text{abs}}} \mathrm{CD}_{\mathrm{co}} = 0$ gives a conventional critical coupling condition:
\begin{equation}
    \gamma_{\text{rad}} \simeq \gamma_{\text{abs}},
\end{equation}
which maximizes the co-polarized circular dichroism.

To illustrate how the dissipation loss determines the rise of transmission CD, we introduce controlled loss in the second MM layer by adding an imaginary component to its permittivity. Figure~\ref{fig:optimal_losses} shows two vertical panels corresponding to Crossing A (left) and Crossing B (right). As the loss increases, we observe a corresponding increase in $\mathrm{CD}_{\text{co}}$, with the same handedness as $\mathrm{CD}_{\text{mode}}$. To illustrate this behavior, we select two thicknesses of the Si layer: $h_1 = 540$~nm and $h_1 = 430$~nm, corresponding to Crossing A [Fig.~\ref{fig:optimal_losses}(a)] and Crossing B [Fig.~\ref{fig:optimal_losses}(b)], respectively. In each plot, the solid and dashed black lines indicate the resonance wavelengths of the modes. The corresponding $\mathrm{CD}_{\text{co}}$ values for these modes are shown in the middle panels [Fig.~\ref{fig:optimal_losses}(c,d)]. The magenta dashed lines indicate a specific value of the imaginary part of the permittivity, at which we plot the circular dichroism to analyze mode interaction, as shown in Fig.~\ref{fig:optimal_losses}(e,f). These results show excellent agreement with the $\mathrm{CD}_{\text{mode}}$ extracted eigenmode spectrum, which is overlaid with the $\mathrm{CD}_{\text{co}}$. It is worth mentioning that $\mathrm{CD}_{\text{mode}}$ is calculated in the absolute absence of dissipation loss in the MM. Therefore, $\mathrm{CD}_{\text{mode}}$ merely predicts the intrinsic capability of a resonant mode to couple preferentially to light of a specific helicity. Manifestation of this coupling selectivity in the form of transmission CD requires additional loss engineering.

\section{Conclusion}
We have studied resonant asymmetric membrane metasurfaces to uncover the physics of their maximum chiral response and engineer chirality.  In contrast to many recently suggested strategies to achieve maximum optical chirality with metasurfaces of complex three dimensional shapes by engineering their geometry in the out-of-plane direction, here we have employed bilayer structuring of rotated C$_4$-symmetric arrays of four-petal holes in dielectric membranes. We have revealed how to maximize optical chirality of such structures by achieving strong coupling between photonic eigenmodes combined with engineering of dissipative losses. Our finding expands substantially the classes of maximally chiral resonant metasurfaces desired for many applications such as chiral sensing and chirality encoding. 

\begin{acknowledgments}
B.K. and I.T. contributed equally to this work. 
The authors thank Alexander Antonov, Kristina Frizyuk, and Peng Xia for useful discussions and suggestions. The work of B.K., I.T. A.K and Y.K has been supported by the Australian Research Council (Grant No. DP210101292) and International Technology Center Indo-Pacific (ITC IPAC) via Army Research Office (Contract FA520923C0023) M.G. acknowledges a support from the Russian Science Foundation  (project 23-42-00091, \url{https://rscf.ru/project/23-42-00091/}).
\end{acknowledgments}

\appendix 
\renewcommand{\thefigure}{\Alph{section}\arabic{figure}}

\section{Co-polarized circular dichroism in lossless rotationally symmetric systems} 
\setcounter{figure}{0}
\label{app:CDzero}

Let us show that the presence of losses is crucial to observe circular dichroism in transmission for the $m$-fold symmetric chiral metasurfaces with $m\geq 3$ under assumption that there is no diffraction. 

The S-matrix of a diffractionless metasurface connects 4 possible input channels amplitudes $a_i$ with 4 possible output channels amplitudes $b_i$: $b_i = \sum_{i=1}^{4}\hat{S}_{ij} a_j$. In the circular basis $\ort{e}_{\pm}$ it is given by~\cite{Gorkunov2020PRL,Shalin2023,Koshelev2024JOPT}
\begin{equation}
    \begin{pmatrix}
        b_+\\b'_+\\b_-\\b'_-
    \end{pmatrix} =
    \begin{pmatrix}
        r_{\text{LR}} &t'_{\text{LL}} & r_{\text{LL}} & t'_{\text{LR}}\\t_{\text{RR}} &r'_{\text{RL}} & t_{\text{RL}} & r'_{\text{RR}}\\r_{\text{RR}} &t'_{\text{RL}} & r_{\text{RL}} & t'_{\text{RR}}\\t_{\text{LR}} &r'_{\text{LL}} & t_{\text{LL}} & r'_{\text{LR}}
    \end{pmatrix}
    \begin{pmatrix}
        a_+\\a'_+\\a_-\\a'_-
    \end{pmatrix}.
    \label{eq:smatrix}
\end{equation}
Here prime indicates excitation from the bottom side of metasurface. First index (R or L) shows the helicity of the outgoing wave and second index (R or L) shows the helicity of ingoing wave, i.e. the pump.
For waves propagating along the $+z$ direction, $\ort{e}_{+}$ ($\ort{e}_{-}$) corresponds to the RCP (LCP) light, while for the $-z$ propagation $\ort{e}_{-}$ ($\ort{e}_{+}$) corresponds to the RCP (LCP) light.
For a metasurface with an arbitrary symmetry transformation matrix $\hat{T}$, the S-matrix must satisfy $\hat{T}^{-1}\hat{S}\hat{T} = \hat{S}$. 
First, we find that Lorentz reciprocity with $(\hat{T}_{\text{R}})_{ij} = \delta_{i,(j+2)\mod 4}$~\footnote{Explicitly transformation matrix reads as \(\hat{T}_{R} = \left(\begin{smallmatrix}
0 & 0 & 1 & 0\\
0 & 0 & 0 & 1\\
1 & 0 & 0 & 0\\
0 & 1 & 0 & 0
\end{smallmatrix}\right)\).} requires that 
$t_{\text{RR}} = t_{\text{RR}}^{\prime}$,
$t_{\text{LL}} = t_{\text{LL}}^{\prime}$,
$t_{\text{RL}} = t_{\text{LR}}^{\prime}$,
$t_{\text{RL}}^{\prime} = t_{\text{LR}}$,
$r_{\text{RL}} = r_{\text{LR}}$, and 
$r_{\text{RL}}^{\prime} = r_{\text{LR}}^{\prime}$. 
Next, $m$-fold rotational symmetry with $\hat{T}_{\phi_m} = \operatorname{diag}(\eu^{\iu \phi_m}, \eu^{\iu \phi_m}, \eu^{-\iu \phi_m}, \eu^{-\iu \phi_m})$, where $\phi_m = 2\pi/m$ results in no additional constrains for $m<3$. Contrary, for  $m\geq 3$ we get
\begin{align}
    r_{\text{LL}} &=r_{\text{LL}}^{\prime}= r_{\text{RR}}  = r_{\text{RR}}^{\prime} = 0, \\
    t_{\text{LR}} &= t_{\text{LR}}^{\prime} = t_{\text{RL}} = t_{\text{RL}}^{\prime} =0.
\end{align}
This means that cross-polarization in transmission is forbidden, and helicity-preserving reflection is also not allowed in this case.
All this combined transforms general form of the S-matrix in Eq.~\eqref{eq:smatrix} to
\begin{equation}
    \hat{S} =  \begin{pmatrix}
    r_{\text{LR}} & t_{\text{LL}} & 0 & 0 \\
    t_{\text{RR}} & r_{\text{RL}}^{\prime} & 0 & 0 \\
    0 & 0 & r_{\text{LR}} & t_{\text{RR}} \\
    0 & 0 & t_{\text{LL}} & r_{\text{RL}}^{\prime}
\end{pmatrix}.
\end{equation}
From here we see four important features of a $C_m$ symmetric metasurface for $m \geq 3$ made of reciprocal materials: 
\begin{enumerate}
    \item[(i)] helicity of light always flips in the reflection ($\mathrm{C}_m$ symmetry); 
    \item[(ii)] cross polarization conversion in transmission is absent ($\mathrm{C}_m$ symmetry); 
    \item[(iii)] it has zero co-polarized circular dichroism in reflection (reciprocity); 
    \item[(iv)] co-polarized circular dichroism in transmission is independent of the excitation side (reciprocity).
\end{enumerate}
Finally, the unitary condition for lossless structures requires that $\hat{S}^{\dagger} \hat{S} = \hat{I}$ which leads to
\begin{equation}
    \begin{split}
        t_{\text{LL}}r^*_{\text{LR}}&=-r'_{\text{RL}}t^*_{\text{RR}}, \\
        t^*_{\text{LL}}r'_{\text{RL}}&=-r^*_{\text{LR}}t_{\text{RR}}.
    \end{split}
    \label{eq:unitary_cond_of_S}
\end{equation}
By substituting $r_{\text{RL}}^{\prime} = - r_{\text{LR}}^{*}  t_{\text{LL}} / t_{\text{RR}}^{*}$ from the first line to the second line we see that $|t_{\text{RR}}|^2 = |t_{\text{LL}}|^2$.
Thus, a lossless reciprocal metasurface with 4-fold rotational symmetry exhibits no co-polarized circular dichroism, 
\begin{equation}
    \mathrm{CD}_{\mathrm{co}} = 0.
    \label{eq:CDcoZERO}
\end{equation}
This result matches with our full numerical simulations done in COMSOL Multiphysics. In Fig.~\ref{fig:optimal_losses}(a,b) we see that $\mathrm{C}_4$ symmetric chiral metasurface gives zero circular dichroism for the lossless case, $\Im(\varepsilon) = 0$. 
On the other hand, extremely high losses are not going to provide a good performance either. Hence, there is an optimal amount of losses which has to be engineered for each particular resonance.

Historically, circular dichroism has been intrinsically linked to asymmetric losses between light of opposite helicities~\cite{Shalin2023,Kobayashi2011}. 
However, we stress that Eq.~\eqref{eq:CDcoZERO} is not a general result, and a non-zero co-polarized circular dichroism can be achieved in lossless or near-lossless systems of other symmetries, such as metasurfaces with monoclinic meta-atom arrangements~\cite{Toftul2024PRL,Toftul2025Nanophotonics} or with $\mathrm{C}_2$ symmetric meta-atoms~\cite{Gryb2023NL,Sinev2024arXiv,Koshelev2024JOPT} or with asymmetric meta-atoms~\cite{Gorkunov2021AOM} to name a few.

\section{Eigenmodes symmetry analyses}
\setcounter{figure}{0}
\label{app:symmetry}

Consider a system that possesses certain symmetries, making it invariant under specific operations such as rotations or inversions. These symmetry operations form a group. We denote by \(\hat{g}\) a symmetry operation which acts any scalar function as
\begin{equation}
    \hat{g} f(\vb{r}) = f(\vb{g}^{-1} \vb{r}),
    \label{eq:gscalar}
\end{equation}
where $\mathbf{g}$ is the transformation matrix which corresponds to this symmetry operation. Once the system is symmetric with respect to this operation we have
\begin{equation}
    \hat{g}\varepsilon(\mathbf{r}) = \varepsilon(\mathbf{r}),
    \quad \hat{g}\mu(\mathbf{r}) = \mu(\mathbf{r}),
\end{equation}
where \(\varepsilon(\mathbf{r})\) and \(\mu(\mathbf{r})\) are the electric permittivity and permeability of the system.
This operator acts on the system's vector field \(\mathbf{A}(\mathbf{r})\) as follows:
\begin{equation}
    \hat{g} \mathbf{A} = \vb{g} \mathbf{A}(\mathbf{g}^{-1} \mathbf{r}),
    \label{eq:gvector}
\end{equation}
where \(\mathbf{g}\) is the transformation matrix. When this rule is applied to a magnetic field, one has to remember that it is a pseudo-vector not a vector. 

Each eigenmode \(\mathbf{E}_n\) of the system transforms into another eigenmode under an irreducible representation~\cite{Hergert2018,Dresselhaus2008,Zee2016,Tsimokha2022PRB}. Specifically, for a group element \(g\), the transformation is given by:
\begin{equation}
    \hat{g} \mathbf{E}_n = \sum_m D_{nm}(g) \mathbf{E}_m,
    \label{eq:general_g}
\end{equation}
where \(D(g)\) represents the representation of the group element \(g\) in the basis of eigenmodes.

\subsection{Properties of a four fold symmetric patterned membrane}

Next, we analyse eigenmodes of the structure which possesses $C_{4h}$ point group symmetry --- a point group symmetry of the unit cell of symmetric membrane. This group, which is a direct product of inversion and 4-fold rotation symmetries, i.e. $C_{4h} = C_4 \otimes i$, has 8 elements: 
\begin{itemize}
    \item[--] Identity: $E$;
    \item[--] Rotations: $C_4$, $C_4^2 = C_2$, $C_4^3$;
    \item[--] Inversion: $i$;
    \item[--] Mirror symmetry: $C_2 \cdot i = \sigma_h$;
    \item[--] Improper rotations: $ C_4 \cdot i = S_4$, $C_4^3 \cdot i  = S_4^3$.
\end{itemize}
Character table is shown in Table~\ref{tab:C4h}. We write explicitly matricies which correspond to all this operations:
\begin{align}
\begin{split}
    \mathbf{g}_{E} &= \begin{pmatrix}
        1 & 0 & 0 \\
        0 & 1 & 0 \\
        0 & 0 & 1
    \end{pmatrix}, \quad
    \mathbf{g}_{C_4} = \begin{pmatrix}
        0 & -1 & 0 \\
        1 & 0 & 0 \\
        0 & 0 & 1
    \end{pmatrix}, \\
    \mathbf{g}_{C_2} &= \begin{pmatrix}
        -1 & 0 & 0 \\
        0 & -1 & 0 \\
        0 & 0 & 1
    \end{pmatrix}, \quad 
    \mathbf{g}_{C_4^3} = \begin{pmatrix}
        0 & 1 & 0 \\
        -1 & 0 & 0 \\
        0 & 0 & 1
    \end{pmatrix}, \\
    \mathbf{g}_{i} &= \begin{pmatrix}
        -1 & 0 & 0 \\
        0 & -1 & 0 \\
        0 & 0 & -1
    \end{pmatrix}, \quad 
    \mathbf{g}_{S_4} = \begin{pmatrix}
        0 & 1 & 0 \\
        -1 & 0 & 0 \\
        0 & 0 & -1
    \end{pmatrix}, \\
    \mathbf{g}_{\sigma_h} &= \begin{pmatrix}
        1 & 0 & 0 \\
        0 & 1 & 0 \\
        0 & 0 & -1
    \end{pmatrix}, \quad 
    \mathbf{g}_{S_4^3} = \begin{pmatrix}
        0 & -1 & 0 \\
        1 & 0 & 0 \\
        0 & 0 & -1
    \end{pmatrix}.
\end{split}
\label{eq:g_C4h}
\end{align}

\begin{table}[h]
\centering
\resizebox{\linewidth}{!}{\begin{tabular}{l|rrrr|rrrr}  %
\hline\hline
$C_{4h} = C_4 \otimes i$ & $E$ & $C_4$ & $C_2$ & $C_4^3$ & $i$ & $iC_4 = S_4$ & $iC_2 = \sigma_h$ & $iC_4^3 = S_4^3$ \\
\hline
$A_g$ & $1$ &  $1$ &  $1$ &  $1$ &  $1$ &  $1$ &  $1$ &  $1$ \\
$B_g$ & $1$ & $-1$ &  $1$ & $-1$ &  $1$ & $-1$ &  $1$ & $-1$ \\
$E_g$ & $2$ &  $0$ & $-2$ &  $0$ &  $2$ &  $0$ & $-2$ &  $0$ \\
$A_u$ & $1$ &  $1$ &  $1$ &  $1$ & $-1$ & $-1$ & $-1$ & $-1$ \\
$B_u$ & $1$ & $-1$ &  $1$ & $-1$ & $-1$ &  $1$ & $-1$ &  $1$ \\
$E_u$ & $2$ &  $0$ & $-2$ &  $0$ & $-2$ &  $0$ & $ 2$ &  $0$ \\
\hline\hline
\end{tabular}}
\caption{Character table of the $C_{4h}$ point group~\cite{Dresselhaus2008,Hergert2018}. Here index $g$ is for symmetric under inversion (gerade) and index $u$ is for antisymmetric under inversion (ungerade).}
\label{tab:C4h}
\end{table}

Depending on how modes transform under group symmetry operations \eqref{eq:g_C4h} there are only 6 possible types. The character value for the identity representation $E$ corresponds to the degeneracy of the modes. 
For non degenerate modes transformation \eqref{eq:general_g} becomes trivial since $D_{nm}$ is going to be scalar value which equal to the character value.

In the far field above ($z\to \infty$) or below ($z\to - \infty$) the structure the wave must be transverse, i.e.
\begin{equation}
    \mathbf{E}_n \simeq \ort{x} E_{n,x} + \ort{y} E_{n,y}.
    \label{eq:En_FF}
\end{equation}
Only modes which transforms under $E_g$ and $E_u$ fulfil this condition, as all others based on Eq.~\eqref{eq:general_g} yeld to $E_{n,x} = -E_{n,x}$ and $E_{n,y} = -E_{n,y}$, hence $E_{n,x} = E_{n,y} = 0$. This means that once eigen mode corresponds to $A_{g,u}$ or $B_{g,u}$, it cannot radiate in the far field, and hence it is a BIC. 
All modes which can radiate, i.e. have a finite $Q$-factor, correspond to either $E_g$ or $E_u$ depending on the up-down parity.
Similar conclusions were made in~\cite{Sadrieva2019PRB,Shalin2023}.

The key operation is rotation by $\pi/2$ about the $z$-axis, which we denote by operator $\hat{g}_{C_4} = \hat{R}_{\pi/2}$. The action of this operator on scaler or vector functions should be understood according to Eqs.~\eqref{eq:gscalar} and~\eqref{eq:gvector}. The explicit form of the $3\times 3$ transformation matrix $\mathbf{R}_{\pi/2} = \mathbf{g}_{C_4}$ is written in Eqs.~\eqref{eq:g_C4h}.

As it is generally implied by Eq.~\eqref{eq:general_g}, a pair of degenerate eigenmodes should transform into their linear superpositions upon $\hat{R}_{\pi/2}$. However, a specific feature of the ${\rm C}_4$ symmetry is that the degenerate modes always transform exactly into each other, which provides a simple form of the $D_{nm}({R}_{\pi/2})$. To prove this, we take an arbitrary representative mode ${\vb{E}}_{1}(\vb{r})$ from a degenerate pair and consider the result of its rotation ${\bf E}'_{1}({\bf r})=\hat{R}_{\pi/2}{\bf E}_1({\bf r})$.
Notably, the scalar product of the 
mode before and after the rotation can be shown to identically vanish as:
\begin{multline}
\label{eq:EprimeE}
     ({\vb E}_1\cdot{\vb E}'_1)=-(\hat{R}^2_{\pi/2}{\vb E}_1\cdot\hat{R}_{\pi/2}{\vb E}_1)\\
     =-(\hat{R}^{-1}_{\pi/2}\hat{R}^2_{\pi/2}{\vb E}_1\cdot{\vb E}_1)=-({\vb E}'_1\cdot{\vb E}_1).
\end{multline}
Here we used the fact that the eigenmodes belongs to the $E$ irrep for which $\hat{R}^2_{\pi/2}{\vb E}_1=\hat{R}_{\pi}{\vb E}_1=-{\vb E}_1$ and assumed the rotation operator to be self-adjoint for the used scalar product definition. 
Then also the norm is preserved upon rotation:
\begin{equation}
\label{eq:EprimeNorm}
     ({\vb E}'_1\cdot{\vb E}'_1)
     =(\hat{R}^{-1}_{\pi/2}\hat{R}_{\pi/2}{\vb E}_1\cdot{\vb E}_1)=({\vb E}_1\cdot{\vb E}_1),
\end{equation}
which shows that the pair ${\vb E}_1$ and ${\vb E}'_1$ constitutes an orthogonal normalized basis in their subspace.

Therefore, we conclude that the mode ${\vb E}'_1$ actually represents the degenerate counterpart of the mode ${\vb E}_1$, which we denote as ${\vb E}_2$ in the main text, and these two modes are related by the operator $\hat{R}_{\pi/2}$ as is described by
\begin{equation}
    D_{R_{\pi/2}} = D_{C_4} = \begin{pmatrix}
        0 & 1 \\
        -1 & 0
    \end{pmatrix},
    \label{eq:DR}
\end{equation}
or Eq.~\eqref{eq:general_g} would read as
\begin{align}
\begin{split}
    \hat{R}_{\pi/2} \vb{E}_1 &= \vb{E}_2 , \\
    \hat{R}_{\pi/2} \vb{E}_2 &= - \vb{E}_1.
\end{split}
\end{align}
The latter is used in the main text. 
Since $D(g_1 g_2) = D(g_1) D(g_2)$, this choice dictates all other matrices $D$ to be
\begin{align}
\begin{split}
    D_{E} &= \begin{pmatrix}
        1 & 0 \\
        0 & 1
    \end{pmatrix}, \quad 
    D_{C_4} = \begin{pmatrix}
        0 & 1 \\
        -1 & 0
    \end{pmatrix}, \quad 
    D_{C_2} = \begin{pmatrix}
        -1 & 0 \\
        0 & -1
    \end{pmatrix}, \\ 
    D_{C_4^3} &= \begin{pmatrix}
        0 & -1 \\
        1 & 0
    \end{pmatrix}, \quad 
    D_{i} = \begin{pmatrix}
        -p & 0 \\
        0 & -p
    \end{pmatrix}, \quad 
    D_{S_4} = \begin{pmatrix}
        0 & -p \\
        p & 0
    \end{pmatrix}, \\
    D_{\sigma_h} &= \begin{pmatrix}
        p & 0 \\
        0 & p
    \end{pmatrix}, \quad 
    D_{S_4^3} = \begin{pmatrix}
        0 & p \\
        -p & 0
    \end{pmatrix}.
\end{split}
\label{eq:D_C4h}
\end{align}
Here $p=\pm 1$ is the mode \textit{parity}.  We assign labels \qquote{\textit{even}} for $p = +1$ and \qquote{\textit{odd}} for $p = -1$. We note that modes in a degenerate pair always have identical parities. One can see that the inversion $i$ parity has opposite sign to the up-down mirror reflection $\sigma_h$ parity for $E_g$ and $E_u$ irreps. 
It is handy to write explicitly the latter one as it has its application in our analyses:
\begin{equation}
    \mathbf{g}_{\sigma_h} \mathbf{E}_n(\mathbf{g}_{\sigma_h}^{-1} \mathbf{r}) = p_n \mathbf{E}_n(\mathbf{r})
\end{equation}
or
\begin{equation}
    \begin{split}
        E_{n\: x,y}(x,y,-z) &= p_n E_{n\: x,y}(x,y,z), \\  
        -E_{n\: z}(x,y,-z) &= p_n E_{n\: z}(x,y,z).
    \end{split}
    \label{eq:parityEnxy}
\end{equation}

Equations \eqref{eq:general_g}, \eqref{eq:g_C4h}, and \eqref{eq:D_C4h} give a full picture how eigen modes which correspond to $E_g$ or $E_u$ transform upon action of any symmetry operations in $C_{4h}$.

\subsection{Consequences of the symmetry analysis for the coupling parameters of a four fold symmetric patterned membrane}

Here we also analyse how properties derived in section~\eqref{app:symmetry} manifest itself in the coupling parameters \eqref{eq:m} and \eqref{eq:m_prime}. 
As a helpful identity, one can check that for any function $f(\vb{r})$ and transformation matrix $\vb{g}$ with $|\det \vb{g}| = 1$ we have 
\begin{equation}
    \int\limits_{V} f(\mathbf{g}^{-1} \mathbf{r}) \dd V = \int\limits_{V^{\prime}} f(\vb{r}^{\prime}) |\det \mathbf{g}| \dd V^{\prime} =  \int\limits_{V} f(\vb{r}) \dd V
    \label{eq:property}
\end{equation}
since volume of integration is invariant upon action of $\mathbf{g}$, $V^{\prime} = V$.

Since there is a scalar product with $\vu{e}_{+}$ or $\vu{e}_{-}$ in Eqs.~\eqref{eq:m}--\eqref{eq:m_prime}, we are interested only in $x$- and $y$-components of the fields.
Plugging property \eqref{eq:parityEnxy}, derived from $\hat{g}_{\sigma_h}$ transformation, into the Eqs.~\eqref{eq:m}--\eqref{eq:m_prime} and using the fact that permittivity and integration volume is invariant upon $z \to -z$,  we find that
\begin{align}
    m_{nR} = p_n m_{nL}^{\prime}, \qquad 
    m_{nL} = p_n m_{nR}^{\prime}. 
    \label{eq:cnsq1}
\end{align}
Based on the $\hat{g}_{C_4}$ transformation we find that
\begin{align}
\begin{split}
    -E_{1y}(y, -x, z) &= E_{\bar{1}x}(x, y, z), \\
    E_{1x}(y, -x, z) &=  E_{\bar{1}y}(x, y, z),
\end{split} \\
\begin{split}
    E_{\bar{1}y}(y, -x, z) &= E_{1x}(x, y, z), \\
    E_{\bar{1}x}(y, -x, z) &= -E_{1y}(x, y, z).
\end{split}
\end{align}
We note that 
\begin{equation}
    [\mathbf{g}_{C_4} \mathbf{E}_n(\mathbf{g}_{C_4}^{-1} \mathbf{r})] \cdot \vu{e}_{\pm} =  \pm \iu  \mathbf{E}_n(\mathbf{g}_{C_4}^{-1} \mathbf{r}) \cdot \vu{e}_{\pm}
\end{equation}
and with the help of Eq.~\eqref{eq:property} this leads to following connection
\begin{align}
\begin{split}
    m_{1R} = - \iu m_{\bar{1}R}, \quad 
    m_{1L} =  \iu m_{\bar{1}L}, \\
    m_{1R}^{\prime} =  \iu m_{\bar{1}R}^{\prime}, \quad 
    m_{1L}^{\prime} = - \iu m_{\bar{1}L}^{\prime}.
\end{split}
\label{eq:cnsq2}
\end{align}

\subsection{Properties of a four fold symmetric patterned double asymmetric double layer membrane}

\begin{figure}
    \centering
    \includegraphics[width=1.0\linewidth]{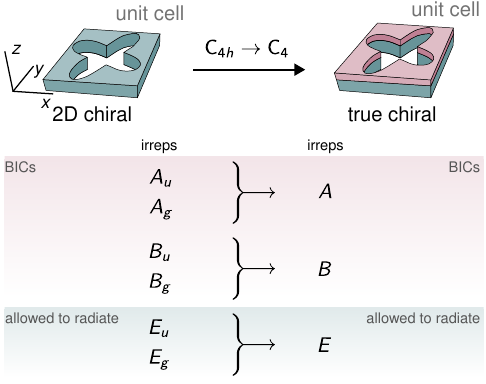}
    \caption{Breaking out-of-plane mirror symmetry transforms a 2D-chiral structure with $C_{4h}$ symmetry into a truly chiral one with $C_4$ symmetry. This transition merges parity-distinguished irreducible representations (irreps) $A_u$ and $A_g$ into $A$, $B_u$ and $B_g$ into $B$, and $E_u$ and $E_g$ into $E$. Modes in $A$ and $B$ remain bound states in the continuum (BICs), while modes in $E$ are allowed to radiate and become relevant for far-field interactions. Only modes from $E$ and $E_{u,g}$ are double degenerate.}
    \label{fig:C4h_to_C4}
\end{figure}

The addition of an adjacent layer break out-of-plane mirror symmetry, i.e. we break the symmetry as $\mathrm{C}_{4h} \to \mathrm{C}_{4}$. Character table of $\mathrm{C}_4$ point group is shown in Table~\ref{tab:C4}. The previously
considered modes which transform  under $A_{u}$ and $A_{g}$  irreps,  lose their specific $\sigma_h$ parity property and now transform under the same irrep $A$. The same occurs to the other modes which transform under different irreps. In Fig.~\ref{fig:C4h_to_C4} we illustrate how different irreps merge into another.

In the present work we are not interested in the modes which transform under $A$ or $B$ irreps as such modes BICs and cannot be coupled to the far field. 
An important consequence of symmetry breaking is that orthogonal modes due to the parity after symmetry break lose its property and they start to interact~\cite{Igoshin2024PRB,CanosValero2024PRRes,Gladyshev2020PRB}.

\begin{table}[h]
\centering
\begin{tabular}{l|rrrr}  %
\hline\hline
$C_{4}$ & $E$ & $C_4$ & $C_2$ & $C_4^3$  \\
\hline
$A$ & $1$ &  $1$ &  $1$ &  $1$   \\
$B$ & $1$ & $-1$ &  $1$ & $-1$  \\  
$E$ & $2$ &  $0$ & $-2$ &  $0$ \\
\hline\hline
\end{tabular}
\caption{Character table of the $C_{4}$ point group~\cite{Dresselhaus2008,Hergert2018}.}
\label{tab:C4}
\end{table}

\section{Coupling parameters of a single layer membrane}\label{app:mRL}
\setcounter{figure}{0}

Plane waves in the asymptotics of eigenmodes of a single layer MM are not necessarily linearly polarized, see e.g. Ref. \cite{Hsu2017}, although controversial statements can be found in earlier works \cite{Zhen2014}. For double degenerate modes of MM considered here, one is always free to choose both of them to be linearly polarized along arbitrary mutually orthogonal directions. In particular, within any degenerate pair, it is easy to transfer to a pair of modes linearly polarized along the coordinate axes. 

One can start with an arbitrary pair of degenerate orthogonal normalized modes ($1,\bar{1}$) and evaluate their coupling parameters $\breve{m}_{1,\bar{2}\ x}$ and $\breve{m}_{1,\bar{1}\ y}$ to $x$ and $y$ polarized waves defined as
\begin{equation}
\begin{split}
    \breve{m}_{n,x} = A_n \int \limits_{V_{\text{MM}}} [\varepsilon(\vb{r}) - 1] \breve{\vb{E}}_n(\vb{r}) \cdot \vu{x} \eu^{-\iu \Omega_n z/c} \dd V, \\
    \breve{m}_{n,y} = A_n \int \limits_{V_{\text{MM}}} [\varepsilon(\vb{r}) - 1] \breve{\vb{E}}_n(\vb{r}) \cdot \vu{y} \eu^{-\iu \Omega_n z/c} \dd V.
\end{split}
\end{equation}
The rotation symmetry conditions \eqref{eq:C4modes} imply that
\begin{equation}\label{eq:m12symm}
    \breve{m}_{1x}=\breve{m}_{\bar{1}y}, \ \ \breve{m}_{\bar{1}x}=-\breve{m}_{1y}.
\end{equation}
Next, one considers a generalized rotation \eqref{eq:U} yielding another pair of orthogonal normalized modes: 
\begin{equation}
\begin{pmatrix}
    \textbf{E}_{1} \\ \textbf{E}_{\bar{1}} 
\end{pmatrix}
= U(\phi) \begin{pmatrix}
    \breve{\textbf{E}}_{1} \\ \breve{\textbf{E}}_{\bar{1}}
\end{pmatrix},
\end{equation}
which are coupled to linearly polarized  waves by the parameters:
\begin{equation}
    \begin{pmatrix}
        {m}_{1x} \\ {m}_{\bar{1}x}
    \end{pmatrix} = U(\phi) 
    \begin{pmatrix}
        \breve{m}_{1x}  \\ \breve{m}_{\bar{1}x}
    \end{pmatrix}, \ 
    \begin{pmatrix}
        {m}_{1y} \\ {m}_{\bar{1}y}
    \end{pmatrix} = U(\phi) 
    \begin{pmatrix}
        \breve{m}_{1y}  \\ \breve{m}_{\bar{1}y}
    \end{pmatrix}.
\end{equation}
Setting the generalized rotation angle to 
\begin{equation}
\phi=-\atan\frac{\breve{m}_{1,y}}{\breve{m}_{\bar{1},y}} 
\end{equation}
eliminates the parameter ${m}_{1y}$, and, by virtue of Eq.~\eqref{eq:m12symm}, also the parameter ${m}_{\bar{1}x}$.

As a result, it is sufficient to introduce a single complex constant $M_1$, and to express all coupling parameters of modes ($1,\bar{1}$) on both metasurface sides as: 
\begin{align}
	m_{1,y} &= m_{\bar{1},x}=m'_{1,y}=m'_{\bar{1},x}=0, \label{eq:m1y} \\ 
	m_{1,x} &= m_{\bar{1},y}=\sqrt{2} M_1, \label{eq:m1x} \\
	m'_{1,x} &= m'_{\bar{1},y}=p_1 \sqrt{2} M_1. \label{eq:m1x1}
\end{align}

For the modes ($1,\bar{1}$) fixed by exactly linearly polarized far-field asymptotics, the other pair of modes ($2,\bar{2}$) cannot be chosen freely, as one has to ensure that the perturbation induces their selective interaction, see Appendix~\ref{app:3pairs}. However, supposing that all modes of achiral membrane possess close-to-linear far-field asymptotics, we can at least approximately express the coupling parameters of modes ($2,\bar{2}$) by another complex constant $M_2$ and a real angle $\psi$ as:
\begin{align}
	m_{2,x} &= m_{\bar{2},y}= \sqrt{2} M_2 \cos\psi,\label{eq:m3x}\\
    m^{\prime}_{2,x} &=m^{\prime}_{\bar{2},y}= p_2 \sqrt{2} M_2 \cos\psi,\label{eq:m3x1}\\
    m_{2,y} &=- m_{\bar{2},x}=\sqrt{2} M_2 \sin\psi,\label{eq:m3y}\\
    m^{\prime}_{2,y} &=-m^{\prime}_{\bar{2},x}= p_2 \sqrt{2} M_2 \sin\psi,\label{eq:m3y1}
\end{align}
where again the rotation symmetry and parity are taken into account

The parameters of coupling of modes to RCP and LCP waves \eqref{eq:m} and \eqref{eq:m_prime} read as: 
\begin{align}
\begin{split}
	m_{n,R}=\frac{m_{n,x}+\iu m_{n,y}}{\sqrt{2}},\  m_{n,L}=\frac{m_{n,x}-\iu m_{n,y}}{\sqrt{2}},\\ 
	m^{\prime}_{n,R}=\frac{m^{\prime}_{n,x}-\iu m^{\prime}_{n,y}}{\sqrt{2}},\  m^{\prime}_{n,L}=\frac{m^{\prime}_{n,x}+ \iu m^{\prime}_{n,y}}{\sqrt{2}}, 
\end{split}
\end{align}
and substituting here Eqs.~(\ref{eq:m1y}--\ref{eq:m3y1}) one obtains the parameters of coupling of the initial modes as:
\begin{align}
\begin{split}
    m_{1,R} &= m_{1,L}=  M_1, \\
    m^{\prime}_{1,R} &= m^{\prime}_{1,L}= p_1 M_1, \\
    m_{\bar{1},R} &= -m_{\bar{1},L}= \iu M_1, \\
    m^{\prime}_{\bar{1},R} &= -m'_{\bar{1},L}=- \iu p_1 M_1, 
\end{split} \\
\begin{split}
    m_{2,R} &=  M_2 \eu^{\iu\psi}, \quad 
    m_{2,L} =  M_2 \eu^{-\iu \psi},\\
    m^{\prime}_{2,R} &= p_2 M_2 \eu^{-\iu \psi}, \quad 
    m^{\prime}_{2,L} = p_2 M_2 \eu^{\iu\psi}, \\
    m_{\bar{2},R} &= \iu M_2 \eu^{\iu\psi}, \quad 
    m_{\bar{2},L} = -\iu M_2 \eu^{-\iu \psi},\\
    m^{\prime}_{\bar{2},R} &= -\iu p_2 M_2 \eu^{-\iu \psi}, \quad 
    m^{\prime}_{\bar{2},L} = \iu p_2 M_2 \eu^{\iu\psi},
\end{split}
\end{align}
which are used as Eqs.~\eqref{eq:M1M2} in the main text.

\section{Transformation of modes by generalized rotation}
\label{app:gen_rot}
\setcounter{figure}{0}

Consider a pair of eigenmodes with field profiles ${\vb E}_1({\vb r})$ and ${\vb E}_2({\vb r})$, which are normalized and mutually orthogonal in accordance with a certain scalar product definition: $({\vb E}_1\cdot{\vb E}_1)=({\vb E}_2\cdot{\vb E}_2)=1$ and $({\vb E}_1\cdot{\vb E}_2)=0$. Particular forms of scalar product can be found in Refs.~\cite{Muljarov2011E,Muljarov2018OL, Lalanne2018,koshelev2022PhDthesis,Gorkunov2025AOM}. Importantly, for the quasi-normal modes, the product is calculated without complex conjugation of the complex field amplitudes.

Then one can introduce a transform that we call \textit{generalized rotation} with arbitrary complex parameter $\varphi$
\begin{eqnarray}
    U(\varphi) = \begin{pmatrix}
        \cos \varphi & \sin \varphi \\
        - \sin \varphi & \cos \varphi
    \end{pmatrix}
    \label{eq:Uapp}
\end{eqnarray}
yielding another pair of modes $\tilde{\vb E}_1({\vb r})$ and $\tilde{\vb E}_2({\vb r})$  according to the rule:
\begin{align}
    \begin{pmatrix}
        \tilde{\textbf{E}}_{1} \\ \tilde{\textbf{E}}_{2}
    \end{pmatrix} = U(\varphi) 
    \begin{pmatrix}
        {\textbf{E}}_{1} \\ {\textbf{E}}_{2}
    \end{pmatrix}.
\end{align}
One can readily check the new modes are also normalized $(\tilde{\vb E}_1\cdot\tilde{\vb E}_1)=(\tilde{\vb E}_2\cdot\tilde{\vb E}_2)=1$ and mutually orthogonal $(\tilde{\vb E}_1\cdot\tilde{\vb E}_2)=0$. Note that this remains true regardless of real or complex character of the angle $\varphi$.

\section{Elements of perturbation matrix}
\setcounter{figure}{0}
\label{app:Vnm}

\begin{figure*}
    \centering
    \includegraphics[width = 0.9\linewidth]{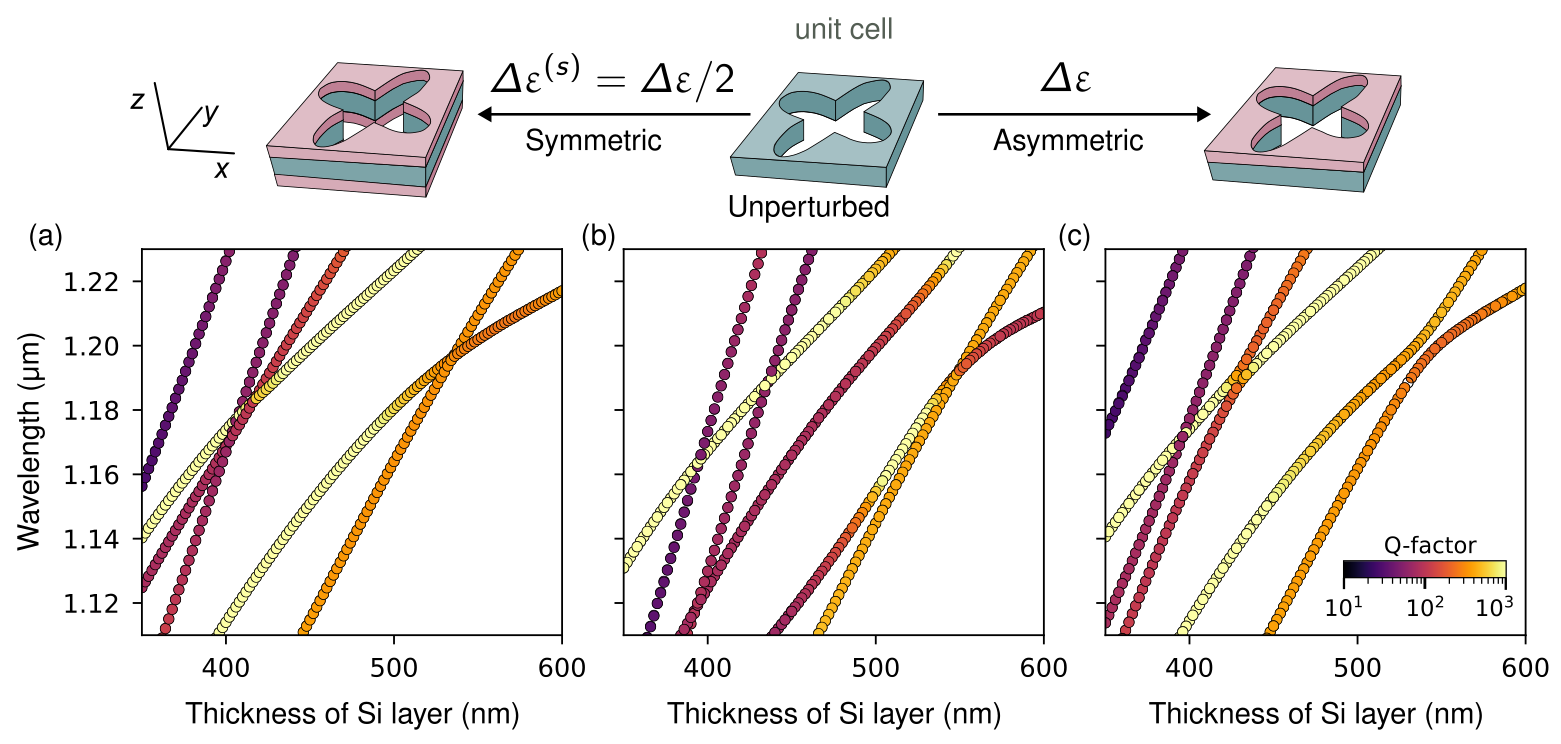}
    \caption{Comparison of MM mode spectra transformations caused by symmetric and asymmetric environment perturbations:
(a) Mode spectrum of MM with symmetric perturbation, where a 200~nm thick second MM layer with permittivity \(\varepsilon_2 = 1+\Delta\varepsilon^s = 2\) is added on both sides. (b) Mode spectrum of the unperturbed single-layer MM. (c) Mode spectrum of MM with with asymmetric perturbation, where a 200 nm-thick MM layer with permittivity \(\varepsilon_2 = 1+\Delta\varepsilon = 3\) (i.e., $\Delta\varepsilon = 2\Delta\varepsilon^s$) is added only on one MM side.}
    \label{fig:sym-Asym perturbation}
\end{figure*}

Generally, the RSE perturbation theory describes the effects of weak changes in the resonating structure properties and shape \cite{Muljarov2011E,Muljarov2018OL}, as well as of its environment  \cite{Almousa2023}. For dielectric systems, all such perturbations are formally described by a variation $\Delta\varepsilon(\textbf{r})$ of the permittivity distribution in space which determines the elements of the perturbation matrix as overlap integrals: 
\begin{equation}
	V_{nm}=\int \Delta\varepsilon(\vb{r}) \varepsilon_0 \vb{E}_n(\vb{r})\cdot\vb{E}_m(\vb{r}) \dd V,
    \label{eq:Vnm}
\end{equation} 
where $\vb{E}_n(\vb{r})$ belong to the normalized orthogonal set of quasinormal modes of unperturbed structure. 

If the perturbation preserves the ${\rm C}_4$ rotation symmetry, it does not couple modes belonging to the same degenerate pair. Indeed, such modes transform into each other as in Eq.~\eqref{eq:C4modes} and for the corresponding matrix elements one can write:
\begin{multline}
    V_{1\bar{1}}=\int \Delta\varepsilon({\bf r}) {\bf E}_1({\bf r})\cdot{\bf E}_{\bar{1}}({\bf r}) \dd V\\ = \int \Delta\varepsilon({\bf r}) {\bf E}_1({\bf r})\cdot\hat{R}_{\pi/2} {\vb{E}}_1(\vb{r}) \dd V\\
    =\int {\bf E}_1({\bf r}) \cdot \hat{R}^{-1}_{\pi/2}\Delta\varepsilon({\bf r})  {\vb{E}}_1(\vb{r}) \dd V\\
    =\int {\bf E}_1({\bf r}) \cdot \Delta\varepsilon({\bf r})  \hat{R}^{-1}_{\pi/2}{\vb{E}}_1(\vb{r}) \dd V\\
    =-\int {\bf E}_1({\bf r}) \cdot \Delta\varepsilon({\bf r})  {\vb{E}}_{\bar{1}}(\vb{r}) \dd V=-V_{1\bar{1}}.\label{eq:V12}
\end{multline}
In the same manner, one can prove the equality of diagonal elements: 
\begin{multline}
    V_{\bar{1}\bar{1}}=\int \Delta\varepsilon({\bf r}) {\bf E}_{\bar{1}}({\bf r})\cdot{\bf E}_{\bar{1}}({\bf r}) \dd V\\ = \int \Delta\varepsilon({\bf r}) {\bf E}_{\bar{1}}({\bf r})\cdot\hat{R}_{\pi/2} {\vb{E}}_1(\vb{r}) \dd V\\
    =\int {\bf E}_1({\bf r}) \cdot \hat{R}^{-1}_{\pi/2}\Delta\varepsilon({\bf r})  {\vb{E}}_{\bar{1}}(\vb{r}) \dd V\\
    =\int {\bf E}_1({\bf r}) \cdot \Delta\varepsilon({\bf r})  \hat{R}^{-1}_{\pi/2}{\vb{E}}_2(\vb{r}) \dd V\\
    =\int {\bf E}_1({\bf r}) \cdot \Delta\varepsilon({\bf r})  {\vb{E}}_1(\vb{r}) \dd V=V_{11}.\label{eq:V22}
\end{multline}

For two pairs of degenerate modes ($1,\bar{1}$) and ($2,\bar{2}$) the transformation of modes upon rotation given by Eq.~\eqref{eq:C4modes} determines a pair of other useful equalities:
\begin{multline}
    V_{\bar{1}\bar{2}}=\int \Delta\varepsilon({\bf r}) {\bf E}_{\bar{1}}({\bf r})\cdot{\bf E}_{\bar{2}}({\bf r}) \dd V\\ = \int \Delta\varepsilon({\bf r}) {\bf E}_{\bar{1}}({\bf r})\cdot\hat{R}_{\pi/2} {\vb{E}}_{2}(\vb{r}) \dd V\\
    =\int {\bf E}_{2}({\bf r}) \cdot \hat{R}^{-1}_{\pi/2}\Delta\varepsilon({\bf r})  {\vb{E}}_{\bar{1}}(\vb{r}) \dd V\\
    =\int {\bf E}_{2}({\bf r}) \cdot \Delta\varepsilon({\bf r})  \hat{R}^{-1}_{\pi/2}{\vb{E}}_{\bar{1}}(\vb{r}) \dd V\\
    =\int {\bf E}_{2}({\bf r}) \cdot \Delta\varepsilon({\bf r})  {\vb{E}}_1(\vb{r}) \dd V=V_{12},\label{eq:V13}
\end{multline}
and 
\begin{multline}
    V_{1\bar{2}}=\int \Delta\varepsilon({\bf r}) {\bf E}_1({\bf r})\cdot{\bf E}_{\bar{2}}({\bf r}) \dd V\\ = \int \Delta\varepsilon({\bf r}) {\bf E}_1({\bf r})\cdot\hat{R}_{\pi/2} {\vb{E}}_2(\vb{r}) \dd V\\
    =\int {\bf E}_2({\bf r}) \cdot \hat{R}^{-1}_{\pi/2}\Delta\varepsilon({\bf r})  {\vb{E}}_1(\vb{r}) \dd V\\
    =\int {\bf E}_2({\bf r}) \cdot \Delta\varepsilon({\bf r})  \hat{R}^{-1}_{\pi/2}{\vb{E}}_1(\vb{r}) \dd V\\
    =-\int {\bf E}_2({\bf r}) \cdot \Delta\varepsilon({\bf r})  {\vb{E}}_{\bar{1}}(\vb{r}) \dd V=-V_{\bar{1}2}.\label{eq:V14}
\end{multline}

For the particular chiral symmetry breaking due to the added dielectric layer the perturbation matrix elements reduce to: 
\begin{equation}
	V_{nm}=\int \limits_0^{h_2}\dd z \int\limits_A \dd x \dd y\ \Delta\varepsilon(x,y) {\bf E}_n({\bf r})\cdot{\bf E}_m({\bf r}), 
    \label{eq:Vnm-our}
\end{equation}
where $A$, as previously, is the metasurface unit cell area.
These overlap integrals couple modes regardless of their parity with respect to $z$-axis inversion, and, as discussed in the Main text, coupling of odd and even modes gives rise to the optical chirality.

To rectify effects of coupling of different modes, it is useful to consider a symmetric perturbation  $\Delta\varepsilon^{(\rm s)}(\textbf{r})$ arising due to adding a pair of identical layers on both MM sides. Notably, the corresponding matrix elements can be expressed as:
\begin{multline}
	V^{(\rm s)}_{nm}=(1+p_np_m)\\
    \times\int\limits_0^{h_2}\dd z \int\limits_A \dd x \dd y \ \Delta\varepsilon^{(\rm s)}(x,y) {\bf E}_n({\bf r})\cdot{\bf E}_m({\bf r}),
    \label{eq:Vnm-sym}
\end{multline}
where, as previously, $p_n=\pm 1$ is a binary indicator of mode parity. 
Accordingly, $V^{(\rm s)}_{nm}=0$ for modes of the opposite parity, and $V^{(\rm s)}_{nm}=V_{nm}$ for modes of the same parity, provided that $\Delta\varepsilon^{(\rm s)}(x,y)=\frac{1}{2}\Delta\varepsilon(x,y)$.  

Exemplary transformations of MM eigenmode spectra induced by such symmetric and antisymmetric perturbations are presented in Fig.~\ref{fig:sym-Asym perturbation}. Comparing subfigures \ref{fig:sym-Asym perturbation}(a) and \ref{fig:sym-Asym perturbation}(c) one can conclude that perturbations of both types induce very similar shifts of the mode eigenfrequency branches. 
Specific differences determined by the perturbation symmetry occur near crossings of the branches. Thus in the region of crossing B odd and even modes freely intersect when the perturbation is symmetric, but exhibit pronounced strong coupling with branch anicrossing when antisymmetric perturbation couples modes of opposite parity. This is fully in-line with the general concept formulated in Ref.~\cite{Gorkunov2025AOM}.

More peculiar situation is observed in the vicinity of crossing A, where branches of three modes, one even and two odd ones, cross when the perturbation is symmetric. This allows concluding that not only the coupling of odd and even modes is absent here, but also two odd modes avoid direct coupling by the symmetric perturbation. Asymmetric perturbation, on the contrary apparently gives rise to the coupling of all modes. Compared to crossing A, however, one can see much weaker Rabi splitting, which indicates that the coupling here is substantially weaker.

\section{Perturbation-induced mixing of 3 pairs of degenerate modes} \label{app:3pairs}
\setcounter{figure}{0}

Consider three pairs of modes (3,$\bar{3}$), (4,$\bar{4}$) and (5,$\bar{5}$) each corresponding to a double degenerate eigenfrequency and transforming into each other within the pairs by rotation as in Eq.~\eqref{eq:C4modes}. In the view of the ambiguity of defining modes withing the pairs, here, at first, these initial modes can be arbitrarily defined. For a specific metastructure perturbation, one can evaluate the corresponding elements of the symmetric perturbation matrix $V_{nm}$, where $n,m=3,\ \bar{3},\ 4,\ \bar{4},\ 5,\ \bar{5}$ as in Eq.~\eqref{eq:Vnm}. For the perturbation retaining the overall rotation symmetry, several elements vanish as in Eq.~\eqref{eq:V12}: $V_{3\bar{3}}=V_{4\bar{4}}=V_{5\bar{5}}=0$. Also the equalities $V_{34}=V_{\bar{3}\bar{4}}$, $V_{35}=V_{\bar{3}\bar{5}}$ and $V_{45}=V_{\bar{4}\bar{5}}$ can be proven as in Eq.~\eqref{eq:V13} as well as the equalities $V_{3\bar{4}}=-V_{\bar{3}4}$, $V_{3\bar{5}}=-V_{\bar{3}5}$ and $V_{4\bar{5}}=-V_{{\bar4}5}$
as in Eq.~\eqref{eq:V14}

For each degenerate pair, one can perform a generalized rotation \eqref{eq:U} yielding another equally valid pair of modes. In particular, for pairs (1,2) and (3,4) one can introduce rotations: 
\begin{align}
    \begin{pmatrix}
        \tilde{\textbf{E}}_{3} \\ \tilde{\textbf{E}}_{\bar{3}}
    \end{pmatrix} &=  U(\phi_3) \begin{pmatrix}
        {\textbf{E}}_{3} \\ {\textbf{E}}_{\bar{3}}
    \end{pmatrix}, \\
    \begin{pmatrix}
        \tilde{\textbf{E}}_{5} \\ \tilde{\textbf{E}}_{\bar{5}}
    \end{pmatrix} &=  U(\phi_5) \begin{pmatrix}
        {\textbf{E}}_{5} \\ {\textbf{E}}_{\bar{5}}
    \end{pmatrix}, 
\end{align}
and evaluate the elements of the perturbation matrix for the rotated states as:
\begin{gather}
    \tilde V_{3\bar{4}}=-\tilde V_{\bar{3}4}=V_{34}\sin\phi_3+V_{3\bar{4}}\cos\phi_3, \label{eq:tV34}\\
    \tilde V_{\bar{4}5}=-\tilde V_{\bar{5}4}=V_{54}\sin\phi_5+V_{5\bar{4}}\cos\phi_5. \label{eq:tV45}
\end{gather}
Choosing the particular values of angles:
\begin{gather}
    \phi_3=-\atan\frac{V_{3\bar{4}}}{V_{34}}, \label{eq:phi3}\\
    \phi_5=-\atan\frac{V_{5\bar{4}}}{V_{54}}, \label{eq:phi5}
\end{gather}
one can ensure that $\tilde V_{3\bar{4}}=\tilde V_{\bar{3}4}=\tilde V_{5\bar{4}}=\tilde V_{\bar{5}4}=0$.

In combination with the empirically established absence of interaction between modes (3,5) and ($\bar{3},\ \bar{5}$), this determines the possibility to write a  mode mixed from all 6 states by the rotation symmetric perturbation as: 
\begin{equation}
\tilde{\textbf{E}}=a_3 \tilde{\bf E}_{3}+a_5 \tilde{\bf E}_{5}+a_4 {\bf E}_{4},\label{eq:E345}
\end{equation}
along with its counterpart: 
\begin{equation}
\tilde{\textbf{E}}'=a_3 \tilde{\bf E}_{\bar{3}}+a_5 \tilde{\bf E}_{\bar{5}}+a_4 {\bf E}_{\bar{4}}.\label{eq:E245}
\end{equation}
These two modes are not only obviously orthogonal, $(\tilde{\textbf{E}}\cdot\tilde{\textbf{E}}')=0$, as they include  mutually orthogonal modes from each degenerate pair, but they also are not coupled by the perturbation. Indeed, such coupling corresponds to the matrix element with several possibly nonzero contributions:
\begin{multline}
    \int \Delta\varepsilon({\bf r}) \tilde{\bf E}({\bf r})\cdot\tilde{\bf E}'({\bf r}) \dd V  = a_3a_5(V_{3\bar{5}}+V_{\bar{3}5})\\
    +a_3a_4(V_{\bar{3}4}+V_{3\bar{4}})+a_5a_4(V_{5\bar{4}}+V_{\bar{5}4})=0,
    \label{eq:VEEprime}
\end{multline}
which all vanish due to the properties of perturbation matrix elements listed above. The possibility to write the mixed state in the form \eqref{eq:E345} is used in the Main text Eq.~\eqref{eq:E123}, where the mixing coefficients $a_{3,4,5}$ are expressed by two mixing angles to preserve the mode normalization during mixing by $a_{3}^2 + a_{4}^2 + a_{5}^2 = 1$.

\section{Fitting of numerical mode branches from RSE equations} \label{app:fitting}
\setcounter{figure}{0}

\begin{figure}
    \centering
    \includegraphics[width=1\linewidth]{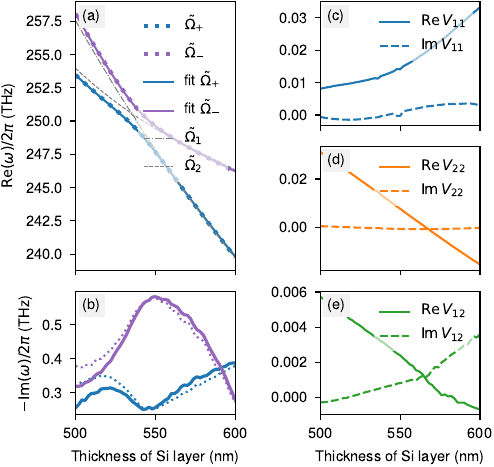}
    \caption{\textbf{Fitting of the crossing A.} (a,b) are identical to the Fig.~\ref{fig:crossingAfit}. (c-e) Values of the $V_{nm}$ coefficients as the functions of the Si lyear thickness. One can see that self-action elements, $V_{11}$ and $V_{22}$, are almost real.}
    \label{fig:crossingAfitExtra}
\end{figure}

\begin{figure}
    \centering
    \includegraphics[width=1\linewidth]{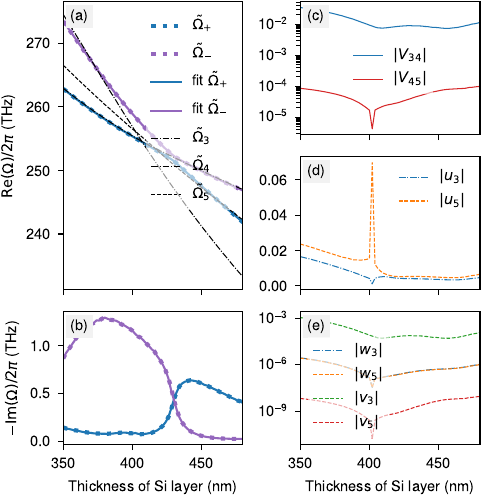}
    \caption{\textbf{Fitting of the crossing B.} (a,b) are identical to the Fig.~\ref{fig:crossingBfit}. (c-e) Absolute values of the $V_{34}$ and $V_{45}$ coefficients, as well as the $u_{3,5}$, $w_{3,5}$, and $v_{3,5}$ as the functions of the Si lyear thickness.}
    \label{fig:crossingBfitExtra}
\end{figure}

To fit the numerically obtained spectrum using the derived analytical model, we first employ the relation between the single-layer eigenfrequencies, $\Omega_n$, and the eigenfrequencies of the symmetric perturbation of equal strength, $\tilde{\Omega}_n$, where the perturbation is described by Eq.~\eqref{eq:Vnm-sym}. This relation provides good estimates of the self-interaction coefficients:
\begin{equation}
    V_{11} \simeq \frac{\Omega_1}{\tilde{\Omega}_1} - 1, \qquad 
    V_{22} \simeq \frac{\Omega_2}{\tilde{\Omega}_2} - 1.
    \label{eq:Vnn}
\end{equation}
Next, the only remaining fitting parameter is the complex coefficient $V_{12}$. For each particular value of the Si layer thickness $h$, we find the optimal value of $V_{12}(h)$ using a standard minimization procedure based on the BFGS method. 
Equation~\eqref{eq:freqsplit} is used as the fitting model, and the numerical values of $\tilde{\Omega}_{\pm}$ obtained from COMSOL are used as target values. 
In order to improve the stability of the fitting procedure and ensure the continuity of $V_{12}(h)$ we use solution of the previous step as the initial guess of the next one. 
The result of this procedure is shown in Fig.~\ref{fig:crossingAfitExtra}. 
Several observations can be made. First, the self-interaction terms are almost real, with negligible imaginary parts: $\Im V_{11} \simeq \Im V_{22} \simeq 0$. Second, the values of all coefficients $V_{nm}$ depend on the thickness of the Si layer. An attempt to fit crossing A within the same parametric range using a constant value of $V_{12}$ resulted in poor agreement for the imaginary parts of the eigenfrequencies, although the real parts were still fitted relatively well.

We implement a similar approach for crossing B, and the result is shown in Fig.~\ref{fig:crossingBfitExtra}. The main difference is that for this case we have two complex fitting parameters, namely $V_{34}$ and $V_{45}$. Self-interaction coefficients $V_{33}$, $V_{44}$, and $V_{55}$ are estimated akin Eq.~\eqref{eq:Vnn}.
One can see a good agreement between the derived model \eqref{eq:freqsplit2} and numerical simulations.

\section{Mode field profiles of a symmetric patterned membrane} 
\setcounter{figure}{0}
\label{app:field_profiles}

In this section we show the spatial field distributions of the eigenmodes involved in Crossings A and B. These modes are computed for a single layer patterned membrane of a $450$~nm thickness.

\begin{figure}
    \centering
    \includegraphics[width=0.95\linewidth]{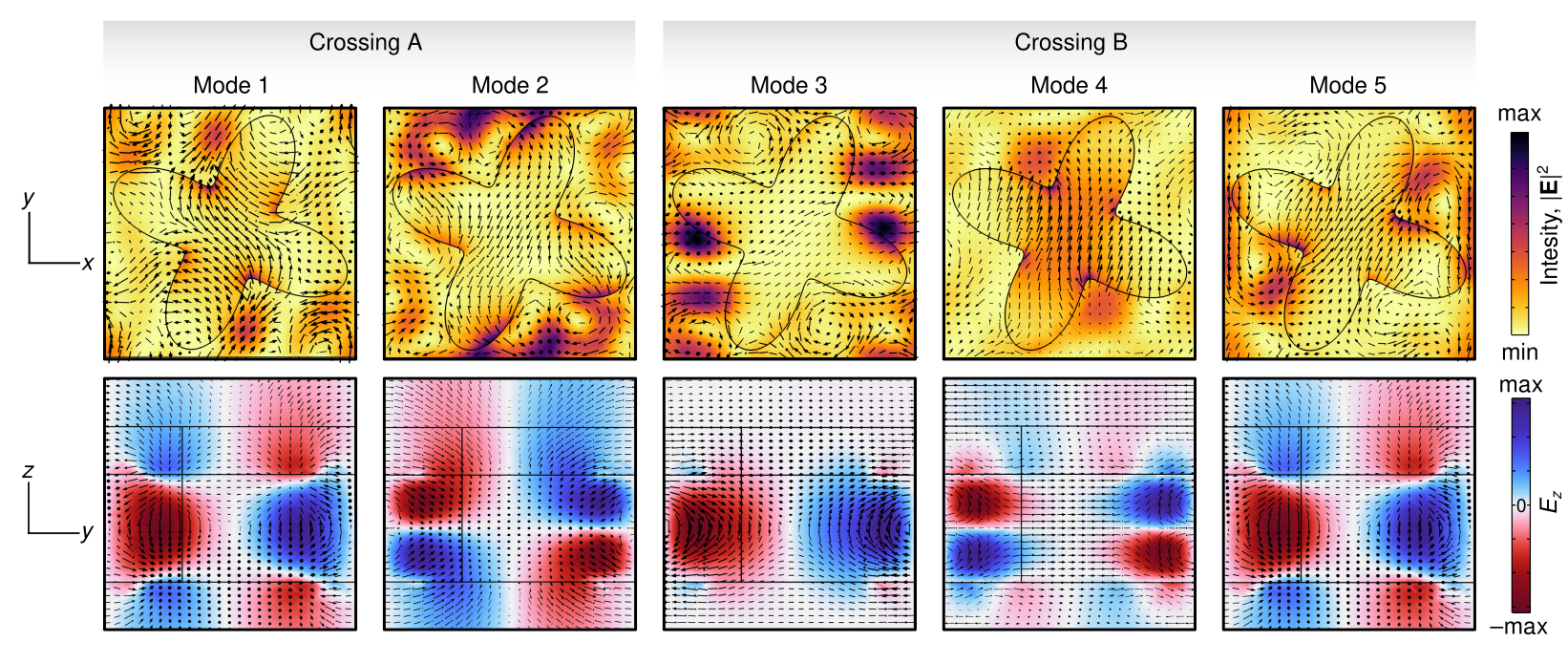}
    \caption{\textbf{Simulated spatial field distributions.} Upper row shows intensity distribution (colormap) of each mode in $xy$-plane, involved in Crossing A (modes 1 and 2), and Crossing B (modes 3, 4, and 5). Lower row represent $E_z$ component of electric field in $yz$-plane, with black arrows indicating electric field vectors. All modes are for a single layer patterned membrane of a $450$~nm thickness.}
    \label{fig:Field_distribution}
\end{figure}

Figure~\ref{fig:Field_distribution} shows five selected modes. The top row displays the electric field intensity  $\propto |\mathbf{E}_n|^2$ in the $xy$-plane placed in the middle of the membrane, overlaid with black arrows representing the electric field vectors. Modes 1 and 2 participate in Crossing A, while modes 3, 4, and 5 are associated with Crossing B. The bottom row shows the $E_z$ component of the electric field in the $yz$-plane, along with corresponding vector fields. 

\onecolumngrid
\ 
\twocolumngrid

\bibliography{refs}

\end{document}